\documentclass[prd,amsfonts,aps,nofootinbib,notitlepage,10pt]{revtex4-1}
\pdfoutput=1
\usepackage{amsmath,amssymb}
\usepackage{graphicx}
\usepackage[utf8]{inputenc}
\usepackage{cancel}
\usepackage[colorlinks=true]{hyperref}
\usepackage{color}
\usepackage[toc,page]{appendix}

\usepackage{slashed}
\usepackage{enumitem}
\newcommand{\nn}{\nonumber}

\begin{document}
%%%%%%%%%%%%%%%%%%%%%%%%%%%%%%%%%%%%%%%%%%%%%%%%%
\title{Inert doublet as multicomponent dark matter}

\author{Amalia Betancur\footnote{\href{mailto:amalia.betancur@eia.edu.co}{amalia.betancur@eia.edu.co}},
  Guillermo Palacio\footnote{\href{mailto:guillermo.palacio38@eia.edu.co}{guillermo.palacio38@eia.edu.co}}\\
\textit{\small  Grupo F\'{i}sica Te\'{o}rica y Aplicada, Universidad EIA,} \\
\textit{\small  A.A. 7516, Envigado, Colombia}\\[4mm]
Andrés Rivera\footnote{\href{mailto:afelipe.rivera@udea.edu.co}{afelipe.rivera@udea.edu.co}}\\
\textit{\small Instituto de F\'{i}sica, Universidad de Antioquia,}\\
\textit{\small Calle 70 No. 52-21, Medell\'{i}n, Colombia}
}

\date{\small \today}

%%%%%%%%%%% ABSTRACT %%%%%%%%%%%%%%%%%%%%%%%%%%%%%%%%%%%%%%
\begin{abstract}
  In this work, we study multicomponent dark sectors comprised of a fermionic and a scalar dark matter candidate. In the scalar sector, we mostly focus on the Inert Doublet Model while in the fermionic sector we study three different models. For all of them, we investigate the impact that
  dark matter conversion and regular WIMP dark matter annihilating processes have on the relic abundance. We mostly recover the region between the electroweak scale and $\sim$ 550 GeV for
  the scalar dark matter mass, which is usually  excluded in the Inert Doublet Model. We also consider current constraints from both direct detection and indirect detection experiments and include future prospects to probe the models. Additionally, we investigate constraints from collider searches on the fermionic dark matter candidates.

\end{abstract}

\maketitle
%%%%%%%%%%%%%%%%%%%%%%%%%%%%  INTRODUCTION %%%%%%%%%%%%%%%%%%%%%%%%%%%%%%%%%
%%%% AMALIA
\section{Introduction}
\label{sec:Introduction}
It is now well established that over 80$\%$ of the total matter content of the Universe is in the form of Dark Matter (DM)~\cite{Aghanim:2018eyx}. Nevertheless, no particle within the Standard Model (SM) of particle physics meets the criteria to be a DM candidate, and so the solution demands physics beyond the SM (BSM). Most models that address the solution include
a weakly interacting massive particle (WIMP), as a DM candidate.
As an example, in the well known Minimal Supersymmetric Standard Model (MSSM), the DM phenomenology focuses usually on the neutralino as the DM candidate, where this neutralino could be Bino, Higgsino, Wino or a mixture of them~\cite{Martin:1997ns}. Some models need far fewer ingredients than the MSSM, for instance, in simple extensions of the SM a field or fields are added such that the lightest neutral particle, if stable, is a DM candidate. In general, the stability requires an additional symmetry which could be a discrete global symmetry such as the $Z_n$ symmetries, with $n=2$ the most widely imposed~\cite{Yaguna:2019cvp, Batell:2010bp, Belanger:2014bga}. Such models tend to be simple, with only a few free parameters, and fractions of them constrained by current experiments; and, because they are economical, they have attracted a great deal of attention. 

One of the most famous simplified models is the Inert Doublet Model (IDM)~\cite{Deshpande:1977rw, LopezHonorez:2006gr} which is a type of Two Higgs Doublet Model (THDM)~\cite{Lee:1973iz, Branco:2011iw}. In this extension, a scalar doublet, similar to the Higgs field, is added to the SM. The field is odd under an imposed $Z_2$ symmetry which renders its lightest neutral component stable and thus, a DM candidate. The popularity of the IDM rests on the fact that it presents an interesting phenomenology for direct detection (DD) \cite{Honorez:2010re,Arina:2009um,Abe:2015rja}, indirect detection (ID) \cite{Andreas:2009hj, Garcia-Cely:2013zga, Garcia-Cely:2015khw, Queiroz:2015utg} and colliders experiments~\cite{Datta:2016nfz,Dutta:2017lny} . Moreover, it has been shown that the IDM may be connected to other BSM problems such as neutrino masses as in the Scotogenic model \cite{Ma:2006km} as well as in the generation of matter and antimatter asymmetry~\cite{Borah:2018rca,Ginzburg:2010wa,Gil:2012ya}. Nevertheless, there are challenges and drawbacks that are worth considering. First and foremost, due to the efficient gauge interactions of the fields, it is only possible to account for the observed relic abundance according to the Planck satellite measurement~\cite{Aghanim:2018eyx} in the Higgs funnel regime ($M_{\rm{DM}}\sim m_h/2$) and $M_{\rm{DM}} \geq 550$ GeV~(
with $M_{\rm{DM}}$ the mass of the DM candidate and $m_h$ the mass of the SM Higgs field). As a result, a region that has great potential from being probed now or in the near future, is not allowed. Moreover, due to the so far null results in WIMP direct DM searches, the viable parameter space is becoming smaller. 

On the other hand, there are no theoretically well-motivated reasons to consider the lightest component of the IDM to be the only DM candidate. As a proof of principle, $5\%$ of the matter-energy of the Universe is composed of a myriad of particles, thus it makes sense to think that the dark sector could be comprised of several stable particles. Models with multicomponent dark sector are gathering attention due to the null results from DM searches~\cite{Zurek:2008qg, Bhattacharya:2016ysw, DiFranzo:2016uzc, Bhattacharya:2018cgx,Karam:2016rsz}. 
Thus, a DM candidate such as the one of the IDM could be accompanied by another stable neutral particle. Works such as \cite{Alves:2016bib, Aoki:2017eqn, Chakraborti:2018aae, Chakraborti:2018lso, Borah:2019aeq,Bhattacharya:2019fgs} have considered the IDM as part of a multicomponent framework where it is accompanied by additional vector boson, fermions, an Axion, and scalar particles respectively.
 
In the present work, we want to investigate the phenomenology of the IDM when it is accompanied by another fermionic weakly interacting massive particle (WIMP) DM candidate. In particular, we want to focus on recovering the scalar DM mass region that goes from $100-550$ GeV although we also consider larger DM masses. To this end, we extend the SM with fermions that are a mix of fields that transform as singlet, doublets, and triplets under the $SU(2)_L$ symmetry. These fields are similar to the well studied Bino-Higgsino, Higgsino-Wino, and Bino-Wino in the MSSM. To stabilize the DM, there are additional global symmetries such that the SM fields are not charged under them, the scalar field is charged only under $Z_2$ while fermionic fields are charged only under the $Z_2'$. For all models, we impose theoretical constraints and investigate the relic density, direct detection, indirect detection and collider experiments restrictions on the parameter space.

This article is organized as follows: In Sec.~\ref{sec:two_DM} we present the formalism for two DM component models, while in Sec.~\ref{sec:scalar_sector} we present a review of the IDM which plays an important role providing the scalar DM candidate, and in Sec.~\ref{sec:pheno_constraints} we discuss the experimental and theoretical constraints applied for all the models proposed.  We also present each of the model's Lagrangian, fields, particle contents with the respective phenomenological analysis and collider constraints  in Sec.~\ref{sec:SD-model} for the singlet-doublet fermion DM  $+$ inert doublet model (SDFDM$+$IDM), in Sec.~\ref{sec:fermionsector}
for doublet-triplet fermion DM  $+$ inert doublet model (DTFDM$+$IDM) and Sec.~\ref{sec:ST-model} for the singlet-triplet fermion dark matter model (STFDM). Finally, we summarize our results in Sec.~\ref{sec:conclusions}.

%%%%%%%%%%%%%%%%%%%%%%%%%%%% ANDRES %%%%%%%%%%%%%%%%%%%%%%%%%%%%%%%%%%%%%
\section{Two DM components}
\label{sec:two_DM}

In this framework, we assume that in the early Universe there are two WIMP particles in the primordial plasma. 
Therefore, we have a multicomponent DM model with two candidates.
Specifically, in this work, the first particle will be the lightest neutral component of an inert scalar~\cite{LopezHonorez:2006gr} and the second one will be a Majorana fermion arising from different representations of the SM's $SU(2)_L$ group, such as a singlet, a doublet or a triplet fermion. This second candidate will be dubbed as $\chi_1^0$ and will emerge in some specific models as we will show latter.

Now, in this general setup of two DM candidates, there are some processes that need to be taken into account in the early Universe in order to  explain the 100\% of the observed DM relic abundance~\cite{Ade:2013zuv}.
First, the DM annihilation $\text{DM}\,\text{DM} \to \text{SMp}\,\text{SMp}$, second the DM semi-annihilation $\text{DM}\,\text{DM} \to \text{DM}\,\text{SMp}$
and finally, the DM conversion~\cite{Belanger:2014vza,Belanger:2018mqt} which involves processes such as   and  $\text{DM}\,\text{DM} \to \text{DM}\,\text{DM}$, where the DM particles can be $H^0$ or $\chi_1^0$ and SMp represents one SM particle.
Nevertheless, in our work, as a result of imposing two discrete symmetries, a $Z_2$ for the scalar DM sector and $Z_2'$ for the fermion DM sector, the  $\text{DM}\,\text{DM} \to \text{DM}\,\text{SMp}$ processes will be forbidden while the other two are still allowed.

To compute the DM relic abundance we used \texttt{MicrOMEGAs}~\cite{Belanger:2014vza,Belanger:2018mqt}. 
This package solves the Boltzmann equations taking into account the last two remaining processes. Those are:
\begin{align}
\dfrac{d\eta_1}{dt}&= - \sigma_v^{11}\left(\eta_1^2 - \bar{\eta}_1^2\right)
-\dfrac{1}{2}\sigma_v^{1122}\left(\eta_1\eta_2 - \eta_2^2\dfrac{\bar{\eta_1}}{\bar{\eta_2}}\right) - 3 H \eta_1 \ , \\
\dfrac{d\eta_2}{dt}&= - \sigma_v^{22}\left(\eta_2^2 - \bar{\eta}_2^2\right)
-\dfrac{1}{2}\sigma_v^{2211}\left(\eta_2\eta_1 - \eta_1^2\dfrac{\bar{\eta_2}}{\bar{\eta_1}}\right) - 3 H \eta_2 \ ,
{\label{eq:boltzman_equations}}
\end{align}
where, $\eta_1 (\eta_2)$ is the number density of the $H^0$($\chi_1^0$) particle, $H$ is the Hubble parameter, $\sigma_v^{ii}$ 
is the thermal averaged cross section for the annihilation process DMi DMi $\to$ SMp SMp (DMi is $H^0$ or $\chi_1^0$) and $\sigma_v^{iijj}$ $(i\neq j)$ 
is the thermal averaged cross section for the conversion process DMi DMi $\to$ DMj DMj.
\begin{figure}[h]
\includegraphics[scale=0.4]{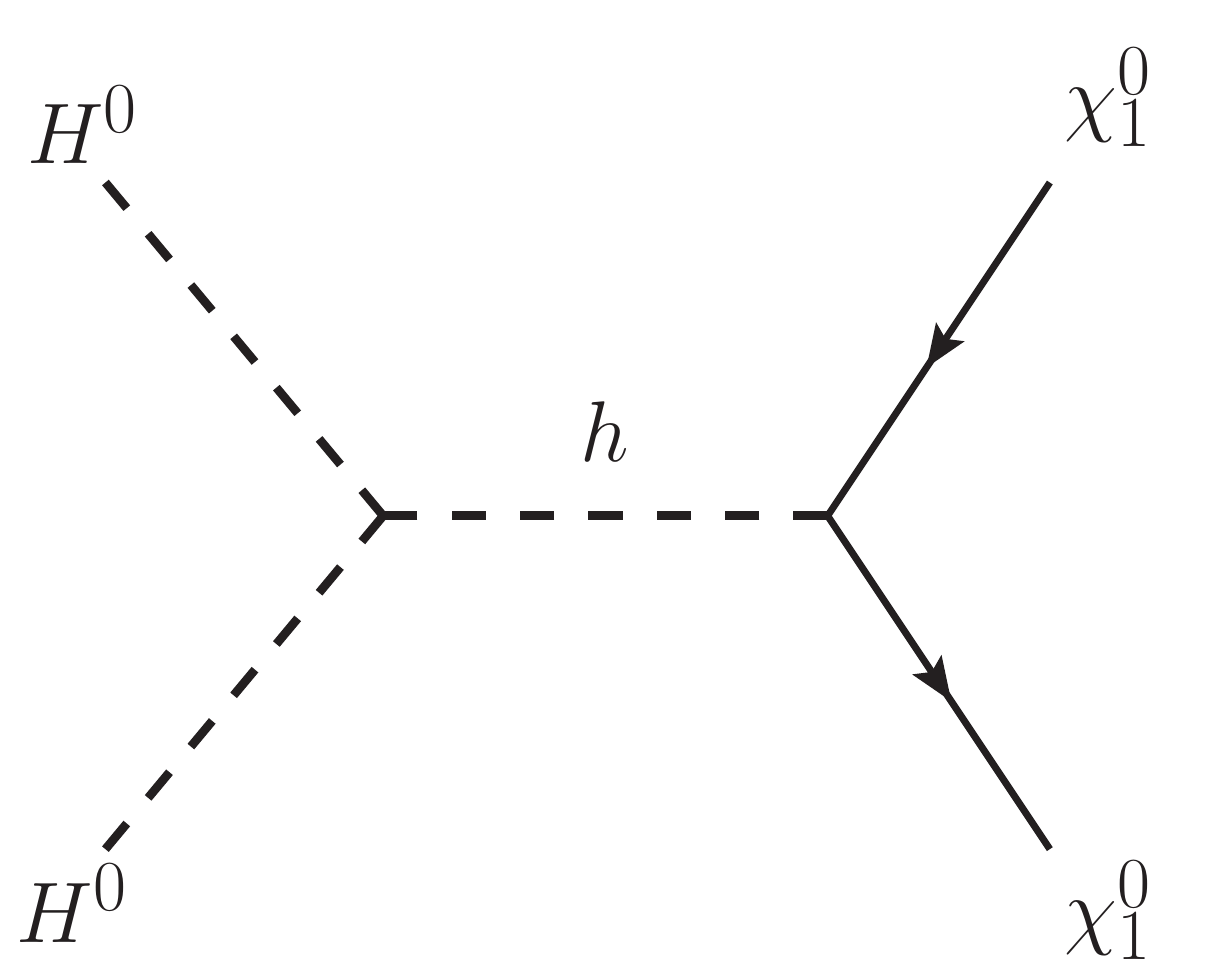}
\caption{DM conversion through the Higgs portal.}
{\label{fig:Higgs_portal}}
\end{figure}
As a result of the nature of this setup and the $Z_2$, $Z_2'$ symmetries, we find that in this work the two DM sectors will communicate only through the Higgs portal as is shown in Fig.~\ref{fig:Higgs_portal}. Nevertheless, \texttt{MicrOMEGAs} takes into account the multiple DM annihilation channels that are natural for each DM model by itself, and so, it includes special processes such as coannihilations and resonances~\cite{Griest:1990kh}.
Therefore, after solving this Boltzmann equations, the program is able to compute the relic abundance for the two DM candidates. 
The contribution from each DM species is displayed, such that
\begin{equation}
\Omega h^2 = (\Omega_{\chi_1^0} h^2 + \Omega_{H^0} h^2)\,.
{\label{eq:sum_DM1_and_DM2}}
\end{equation}
%%%%%%%%%%%%%%%%%%%%%%%% IDM model %%%%%%%%%%%%%%%%%%%%%%%
\section{IDM}
\label{sec:scalar_sector}
The IDM enlarges the SM with an extra scalar doublet, where the new field is odd under a $Z_2$ symmetry, whereas all the other fields are even ~\cite{Deshpande:1977rw,Barbieri:2006dq}. The corresponding scalar potential takes the form
\begin{align}
V(H,\eta)&= -\mu_{1}^2 |H|^2 + \frac{\lambda_1}{2} |H|^4 + \mu_{2}^2 |\eta|^2 + \frac{\lambda_2}{2} |\eta|^4+\lambda_3|H|^2|\eta|^2 + \lambda_4|H^{\dagger} \eta|^2  + \frac{\lambda_5}{2} \left[(H^{\dagger} \eta)^2 + \rm{h.c.} \right],
\end{align}
where the $H$ stands for the Higgs doublet and $\eta$ is the $Z_2$-odd scalar field, which are expressed as
\begin{align}
  \label{eq:scalarcontent}
H=\left( \begin{array}{ccc}
G^+  \\
\frac{v + h +i G^0 }{\sqrt{2}} \end{array} \right),\hspace{2cm}
\eta=\left( \begin{array}{ccc}
H^+  \\
\frac{H^0+i A^0 }{\sqrt{2}} \end{array} \right)~.
\end{align}

After electroweak symmetry breaking (EWSB),
the Higgs field develops a vacuum expectation value (VEV) $ \langle H \rangle = (0,  \frac{v}{\sqrt{2}} )^{T}$, with $v=246$~GeV.
$G^{+}$ and $G^{0}$ becomes
the longitudinal degrees of freedom of $W$ and $Z$ respectively.
Due
 to the quartic couplings to the Higgs,
 the particles within $Z_2$-odd doublet acquire  masses which are given by:

\begin{align}
M_{H0}^2&=\mu^2_{2} + \frac{(\lambda_3 + \lambda_4 +\lambda_5)}{2}v^2 , \\
M_{A0}^2&=\mu^2_{2} + \frac{(\lambda_3 + \lambda_4 -\lambda_5)}{2}v^2 ,  \\
M_{H^+}^2&=\mu^2_{2} + \frac{\lambda_3v^2 }{2}. 
\end{align}

The particle content of IDM (after EWSB) will become part of the scalar sector of the two component DM models  that
we are going to explore, for that reason,
in order to do a complete analysis of the these models, we carry out a scan of the IDM's parameter space as is shown in Table~\ref{table:IDM_scan}.
\begin{table}[h]
\centering
\begin{tabular}{|c|c|} 
\hline
Parameter & Range\\
\hline
$\lambda_{2,3,4}$ & $10^{-4}-10$ \\
$-\lambda_{5}$ & $10^{-8}-10$ \\
$\mu_2$  & $10 - 5\times10^{3}$ (GeV) \\
\hline
\end{tabular}
\caption{Scan range of the parameters of the IDM model.}
\label{table:IDM_scan}
\end{table}
The IDM may be probed by DD experiments, its SI cross section is given by: 
\begin{align}
\label{eq:IDM_SI}
\sigma_{SI}^{H^{0}} = \dfrac{m_r^2}{\pi} \left(\dfrac{\lambda_L}{ m_{H^0} m_h^2}\right)^2f_N^2 m_N^2,
\end{align}
where, $f_N \approx 0.3 $ is the form factor for the scalar interaction~\cite{Alarcon:2011zs,Alarcon:2012nr},
  $M_N \approx 0.938~\rm{GeV}$ is the nucleon mass,
  $m_{r}$ is the reduced mass of the DM and the nucleon defined as
  $m_{r}=M_N m_{H^0}/ (M_N + m_{H^0} )$ and  $\lambda_L=\dfrac{\lambda_3+\lambda_4+\lambda_5}{2}$.

%%%%%%%%%%%%%%%% Theoretical & Collider constraints %%%%%%%%%%%%%%%%%%%%%%%%%%%
\section{Experimental and theoretical constraints}
\label{sec:pheno_constraints}
%\input{Theo_constraints}
%%%%%%%%%%%%%%%%%%%%%%%%%%%%%%%%%%%%%%
%Experimental constraints.
In this section we list experimental and theoretical constraints that will be imposed in all models: 
\begin{enumerate}[label=\roman*]
\item ) Electroweak precision observables (EWPO): Physics BSM can generate  changes on SM observables that arise through loop corrections.
 The set of EWPOs are minimally described by the $STU$
 Peskin-Takauchi parameters~\cite{Peskin:1991sw}.
The $S$ and $T$ oblique parameters are  defined in the standard parametrization
as~\footnote{The U parameter is not displayed since it turns to be small for the three BSM models
under consideration.}~\cite{Peskin:1991sw}:
%%%%%%%%%%%%%%%%%%%%%%%%%%%%%%%%%%%%%%%%%%%%%%%%%%%%%%%%%%%%%%%%%%%%%%%%%%%%%%%%%%%%%%%%%%%%%%%%%%%%%%%%%%%%
\begin{align}
  &S=\dfrac{4s_W^{2}c_W^{2} }{\alpha} \Bigg ( \dfrac{\Pi_{ZZ}(m_{Z}^{2}) - \Pi_{ZZ}(0) }{m_{Z}^{2}}
    - \dfrac{c_{W}^{2} - s_{W}^{2} }{s_W c_W}  \dfrac{\Pi_{Z\gamma}(m_{Z}^{2}) }{m_{Z}^{2}}
    - \dfrac{\Pi_{\gamma \gamma}(m_{Z}^{2}) }{m_{Z}^{2}} \Bigg ) {\label{eq:theo_def_STU-a}},  \\
 &T=\dfrac{1}{\alpha} \Bigg ( \dfrac{\Pi_{W W}(0) }{m_{W}^{2}} - \dfrac{\Pi_{Z Z}(0) }{m_{Z}^{2}} \Bigg) {\label{eq:theo_def_STU-b}},
\end{align}  
%%%%%%%%%%%%%%%%%%%%%%%%%%%%%%%%%%%%%%%%%%%%%%%%%%%%%%%%%%%%%%%%%%%%%%%%%%%%%%%%%%%%%%%%%%%%%%%%%%%%%%%%%%%%
with $\Pi_{V V^{\prime}}$\footnote{ where $ {V V^{\prime}} \in \lbrace W, Z, \gamma \rbrace  $.}
the gauge boson self-energy functions. The new  particle content
of the two component DM models proposed in this work will contribute to the  $\Pi_{V V^{\prime}}$.
We demand that all models fulfill the current experimental limits on S and T~\cite{Tanabashi:2018oca}
%%%%%%%%%%%%%%%%%%%%%%%%%%%%%%%%%%%%%%%%%%%%%%%%%%%%%%%%%%%%%%%%%%%%%%%%%%%%%%%%%%%%%%%%%%%%%%%%%%%%%%%%%%%%
\begin{align}
  &S= 0.02 \pm 0.10~,  {\label{eq:exper_STU-a}} \\
  &T=  0.07 \pm 0.12~. {\label{eq:exper_STU-b}}    
  %&U=  0.00 \pm 0.09 {\label{eq:exper_STU-c}}
\end{align}  
%%%%%%%%%%%%%%%%%%%%%%%%%%%%%%%%%%%%%%%%%%%%%%%%%%%%%%%%%%%%%%%%%%%%%%%%%%%%%%%%%%%%%%%%%%%%%%%%%%%%%%%%%%%%
\item ) From Planck satellite measurements, the DM relic abundance is constrained to be~\cite{Aghanim:2018eyx}:
%%%%%%%%%%%%%%%%%%%%%%%%%%%%%%%%%%%%%%%%%%%%%%%%%%%%%%%%%%%%%%%%%%%%%%%%%%%%%%%%%%%%%%%%%%%%%%%%%%%%%%%%%%%%
\begin{align}
  \Omega h^{2}= 0.1200 \pm 0.0012~.{\label{eq:exper_relic_abundance}}
\end{align}  
%%%%%%%%%%%%%%%%%%%%%%%%%%%%%%%%%%%%%%%%%%%%%%%%%%%%%%%%%%%%%%%%%%%%%%%%%%%%%%%%%%%%%%%%%%%%%%%%%%%%%%%%%%%%
\item ) Additional charged particles may contribute to the branching ratio of the Higgs into two photons. In this work, both the fermionic and scalar dark sectors, should, in principle, contribute to the Higgs diphoton decay rate due to the new charged particles. For the SDFDM+IDM and the STFDM such contribution is not possible due to the $SU(2)_L$ symmetry but for the DTFDM+IDM it is present and may be sizable. Moreover, the charged particles of the scalar sector contributes for all models, though this contribution is usuaylly not sizable. To ensure the correct decay rate we computed its value for all models and ensured that the limit on the signal strength relative to the standard model prediction was always within the allowed values presented by the CMS~\cite{Sirunyan:2018ouh} and ATLAS~\cite{Aaboud:2018xdt} experiments, that is, 
$R_{\gamma \gamma} = \dfrac{Br(h \to \gamma \gamma)_{\text{Observed}}}{ Br(h \to \gamma \gamma)_{SM}}$ are:
%%%%%%%%%%%%%%%%%%%%%%%%%%%%%%%%%%%%%%%%%%%%%%%%%%%%%%%%%%%%%%%%%%%%%%%%%%%%%%%%%%%%%%%%%%%%%%%%%%%%%%%%%%%%
\begin{align}
  R_{\gamma \gamma}^{\rm{CMS}} = 1.18^{+0.17}_{-0.14}~, {\label{eq:Diphoton_rate-a}} \\
  R_{\gamma \gamma}^{\rm{ATLAS}} = 0.99^{+0.15}_{-0.14}~. {\label{eq:Diphoton_rate-b}}
\end{align}  
%%%%%%%%%%%%%%%%%%%%%%%%%%%%%%%%%%%%%%%%%%%%%%%%%%%%%%%%%%%%%%%%%%%%%%%%%%%%%%%%%%%%%%%%%%%%%%%%%%%%%%%%%%%%
\item ) In the SDFDM (in Sec.~\ref{sec:SD-model})
and DTFDM (in Sec.~\ref{sec:fermionsector}) models,
the scalar sector of the SM
is extended introducing a scalar inert doublet (see Sec.~\ref{sec:scalar_sector} ).
 The models are subject to theoretical restrictions
such as perturbativity, vacuum stability and unitarity.
These conditions imply that there are restrictions
for the $\lambda_i$ couplings as well as restrictions
among the couplings themselves as follows \cite{Arhrib:2012ia, Arhrib:2013ela, Belyaev:2016lok}. 
For vacuum stability this is:
\begin{align}
\label{eq:conditionVpositive}
 & \lambda_{1,2}>0\,, \hspace{2cm} \lambda_3 + 2 \sqrt{\lambda_1\lambda_2}>0\,, \hspace{2cm} \; \lambda_3 + \lambda_4 - |\lambda_5| + 2 \sqrt{\lambda_1\lambda_2} >0\,.
\end{align}

For perturbativity, all dimensionless couplings on
the scalar potential must satisfy:

\begin{align}
\label{eq:perturbativityScalar}
|\lambda_i|< 8 \pi\,.
\end{align}

For unitarity we have \cite{Arhrib:2012ia, Arhrib:2013ela, Belyaev:2016lok}:
\begin{align}
|e_i| \leq 8 \pi\,,
\end{align}
where $e_i$ are:
\begin{align}
\label{eq:unitarity}
& e_{1,2}=\lambda_3 \pm \lambda_4,  \  e_{3,4}=\lambda_3 \pm \lambda_5\,,
\nonumber\\ 
& e_{5,6}=\lambda_3 + 2\lambda_4 \pm 3\lambda_5,  \ e_{7,8}=-\lambda_1-\lambda_2 \pm \sqrt{(\lambda_1-\lambda_2)^2 + \lambda_4^2}\,,
\nonumber\\
& e_{9,10}=-3\lambda_1-3\lambda_2 \pm \sqrt{9(\lambda_1-\lambda_2)^2 + (2\lambda_3 +\lambda_4)^2}\,,
\nonumber\\
& e_{11,12}=-\lambda_1-\lambda_2 \pm \sqrt{(\lambda_1-\lambda_2)^2 + \lambda_5^2}\,.
\end{align}

Since the scalar content and potential parameter of the STFDM in Sec.~\ref{sec:ST-model} is different than the one of the two models  mentioned above, we considered the limits used in Ref.~\cite{Merle:2016scw}
%%%%%%%%%%%%%%%%%%%%%%%%%%%%%%%%%%%%%%%%%%%%%%%%%%%%%%%%%%%%%%%%%%%%%%%%%%%%%%%%%%%%%%%%%%%%%%%%%%%%%%%%%%%%
\begin{align}
  \lambda_{1,2} \geq 0~, \hspace{4.7cm}  \lambda_{2}^{\Omega} \geq 0~,  {\label{eq:kk_ll-a}} \\
  \lambda_{3} + \sqrt{\lambda_1 \lambda_2} \geq 0~, \hspace{3cm}  \lambda_{345} + \sqrt{\lambda_1 \lambda_2} \geq 0~, {\label{eq:kk_ll-b}} \\
  \lambda_{1}^{\Omega} + \sqrt{2\lambda_1 \lambda_2^{\Omega}} \geq 0~, \hspace{3cm}  \lambda^{\eta} + \sqrt{2\lambda_2 \lambda_2^{\Omega}} \geq 0~, {\label{eq:kk_ll-c}} 
\end{align}  
%%%%%%%%%%%%%%%%%%%%%%%%%%%%%%%%%%%%%%%%%%%%%%%%%%%%%%%%%%%%%%%%%%%%%%%%%%%%%%%%%%%%%%%%%%%%%%%%%%%%%%%%%%%%
and
%%%%%%%%%%%%%%%%%%%%%%%%%%%%%%%%%%%%%%%%%%%%%%%%%%%%%%%%%%%%%%%%%%%%%%%%%%%%%%%%%%%%%%%%%%%%%%%%%%%%%%%%%%%%
\begin{align}
  \sqrt{2 \lambda_1 \lambda_2 \lambda_{2}^{\Omega} } +
  \lambda_3 \sqrt{2 \lambda_{2}^{\Omega}}  +
  \lambda_{1}^{\Omega}  \sqrt{\lambda_{2} } +
  \lambda^{\eta} \sqrt{\lambda_{1}} +
  \sqrt{\Big (  \lambda_3  + \sqrt{ \lambda_1 \lambda_2 } \Big )
    \Big (  \lambda_{1}^{\Omega}  + \sqrt{2 \lambda_1 \lambda_{2}^{\Omega} } \Big)
    \Big (  \lambda^{\eta}  + \sqrt{2 \lambda_2 \lambda_{2}^{\Omega} } \Big )}
  \geq 0\,, \label{eq:stfdm_bfb_2}
\end{align}
%%%%%%%%%%%%%%%%%%%%%%%%%%%%%%%%%%%%%%%%%%%%%%%%%%%%%%%%%%%%%%%%%%%%%%%%%%%%%%%%%%%%%%%%%%%%%%%%%%%%%%%%%%%%
where $\lambda_{345}= \lambda_3 + \lambda_4 - |\lambda_5| $.
When $\lambda_4 + |\lambda_5| < 0$, in the
equations~\eqref{eq:kk_ll-b} and~\eqref{eq:stfdm_bfb_2}
we must replace $\lambda_3 \to \lambda_{345}$.

\item ) Finally, LEP sets limits on the
masses of charged and neutral particles which couples to the $Z$ and $W$ bosons. The constraints are summarized as~\cite{Lundstrom:2008ai}:

\begin{align}
 &m_{\rho^{\pm}}>103.5~{\rm{GeV}}~,&  &m_{\phi_1^{0}} + m_{\phi_2^{0}} > m_Z~,& &2m_{\rho^{+}} > m_Z~,&       \\
 &m_{\phi^{\pm}}>70.0~{\rm{GeV}}~,&  &2m_{\phi^{\pm}} > m_Z~,&   &m_{\phi_1^{0}} + m_{\phi^{\pm}} > m_W~,&   \\
 &{\rm{max}}(m_{\phi_1^{0}}~, m_{\phi_2^{0}})>100.0~{\rm{GeV}}~,& &m_{\rho^{0}} + m_{\rho^{+}} > m_W~,&   &m_{\phi_2^{0}} + m_{\phi^{\pm}} > m_W~,&   \label{eq:lep_limits}
\end{align}  

where the $\phi$ ($\chi$) stand for scalar (fermions) particles. The particles per model
are displayed in table~\ref{table:fields_at_LEP_constraints}.

\begin{table}[h!]
\centering
\begin{tabular}{|c|c|c|c|c|} 
\hline
 Models /Fields & $\rho^{\pm}$ &  $\rho^{0}$ & $\phi^{\pm}$  & $\phi_i^{0}$   \\
\hline
SDFDM $+$ IDM  & $\chi^{\pm}$ &  ${ \chi_{1,2,3}^{0}}$ & $H^{\pm}$   & ${ H^{0} , A^{0} }$ \\
\hline
DTFDM + IDM  & $\chi_{1,2}^{\pm}$ & ${ \chi_{1,2,3}^{0}}$ & $H^{\pm}$   &  ${ H^{0} , A^{0} }$ \\
\hline
STFDM & $\chi^{\pm}$ & ${ \chi_{1,2}^{0}}$ & $\eta^{\pm}$   &  ${ \eta^{R} , \eta^{I} }$ \\
\hline

\end{tabular}
\caption{Fields appearing in the LEP constraints for the three models under consideration. }
\label{table:fields_at_LEP_constraints}
\end{table}

\end{enumerate}

\section{singlet-doublet fermion dark matter model}
\label{sec:SD-model}

The singlet-doublet DM model, dubbed as the SDFDM for short, has been widely studied in the Ref.~\cite{Cohen:2011ec,Cheung:2013dua,DEramo:2007anh,Abe:2014gua,Calibbi:2015nha,Restrepo:2015ura,Horiuchi:2016tqw}. The model has a rich phenomenology, with possible signals of DD and ID that can be tested in  experiments such as XENON1T~\cite{Aprile:2018dbl}, 
%PandaX~\cite{Cui:2017nnn}, LZ~\cite{Akerib:2018lyp}, 
DARWIN~\cite{Aalbers:2016jon}, Fermi-LAT~\cite{Ackermann:2015zua}, H.E.E.S.~\cite{Abdallah:2018qtu} , etc. 
The SDFDM can also generete neutrino masses at one-loop level if the scalar content of the model is extended as shown in Ref.~\cite{Restrepo:2015ura}.

The particle content of the model consists of one vector-like Dirac $SU(2)_L$-doublet fermion $\Psi=(\Psi^0, \Psi^-)$  and one Majorana singlet fermion $N$ with zero hypercharge, all of them are odd under the $Z_2'$ symmetry, under which the SM particles are even. 
The most general $Z_2'$-invariant Lagrangian includes:
\begin{align}
\label{eq:lt13a}
 \mathcal{L}\supset & \,-M_\Psi \overline{\Psi}\Psi -M_N \overline{N^c}N
 -\left[y_1\, \overline{\Psi}\widetilde{H}P_R N
 +y_2 \overline{\Psi}\widetilde{H}P_L N + \text{h.c.}\right],
\end{align}
where $H$ is the SM Higgs doublet with $\widetilde{H}=i\sigma_2H^{*}$ and $P_{R,L}=(1\pm \gamma_5)/2$ are the projection operators.

After, EWSB the $Z_2'$-odd fermion spectrum is composed by a
charged Dirac fermion $\chi^-$ with a mass $m_{\chi^\pm}=M_\Psi$, and three Majorana fermions that arise from
the mixture between the neutral parts of the $SU(2)_L$ doublets and
the singlet fermion. 
In the basis $\left(N,\Psi_L^0,\left( \Psi_R^0 \right)^{\dagger}
\right)^T$,
the neutral fermion mass matrix is given by:
\begin{align}
\label{eq:Mchi}
  \mathbf{M}=\begin{pmatrix}
 M_N                 &-m_{\lambda}\cos\beta&m_{\lambda}\sin\beta\\
-m_{\lambda}\cos\beta &  0                  & M_\Psi\\
m_{\lambda}\sin\beta&  M_\Psi                &  0  \\
\end{pmatrix},
\end{align}
where
\begin{align}
  \label{eq:etabeta}
\lambda=&\sqrt{y_2^2+y_1^2}\,,&  
m_{\lambda}=&  \frac{\lambda v }{\sqrt{2}}\,,&
  \tan\beta=&\frac{y_2}{y_1}\,.
\end{align}
Note that the mass matrix $ \mathbf{M}$ follows the same
convention of the bino-higgsino sector of the MSSM~\cite{Martin:2012us} 
where $m_\lambda=m_Z\sin\theta_W$ ($\lambda=g'/\sqrt{2}$).
The Majorana fermion mass eigenstates $\mathbf{X}=(\chi_1,\chi_2,\chi_3)^T$ are obtained through the rotation matrix $\mathbf{O}$, such that
%\begin{align}
%\label{eq:chidiag}
$\mathbf{O}^{\operatorname{T}}\mathbf{M} \mathbf{O}=\mathbf{M}^\chi_\text{diag}$,
%\end{align}
with
$\textbf{M}^\chi_{\text{diag}}=\operatorname{Diag}(m_{\chi_1},m_{\chi_2},m_{\chi_3})$. The lightest $\chi_i$ eigenstate will the DM particle and will be dubbed as $\chi_1^0$.
Moreover, in the limit of small doublet-fermion mixing ($m_\lambda\ll M_D,M_N)$, these fermion masses were computed in  Ref.~\cite{Restrepo:2015ura}.

As we mentioned in Sec.~\ref{sec:pheno_constraints}, the new fermions in this model affects the EWPO parameters. The contribution to the $S$ and $T$ parameters were computed in  Ref.~\cite{D'Eramo:2007ga, Abe:2014gua}. We took this into account in the numerical analysis of the SDFDM model, and we used the restriction shown in Sec.~\ref{sec:pheno_constraints}, eqs.~\eqref{eq:exper_STU-a} and~\eqref{eq:exper_STU-b}. 
On the other hand, we also computed numerically the branching ratio of the Higgs decay into two photons and we took into account the current experimental limits of CMS and ATLAS described in Sec.~\ref{sec:pheno_constraints}, eqs.~\eqref{eq:Diphoton_rate-a} and~\eqref{eq:Diphoton_rate-b}.

\subsection{DM conversion example}
\label{sec:SDFDM-conversion}

\begin{figure}
\includegraphics[scale=0.4]{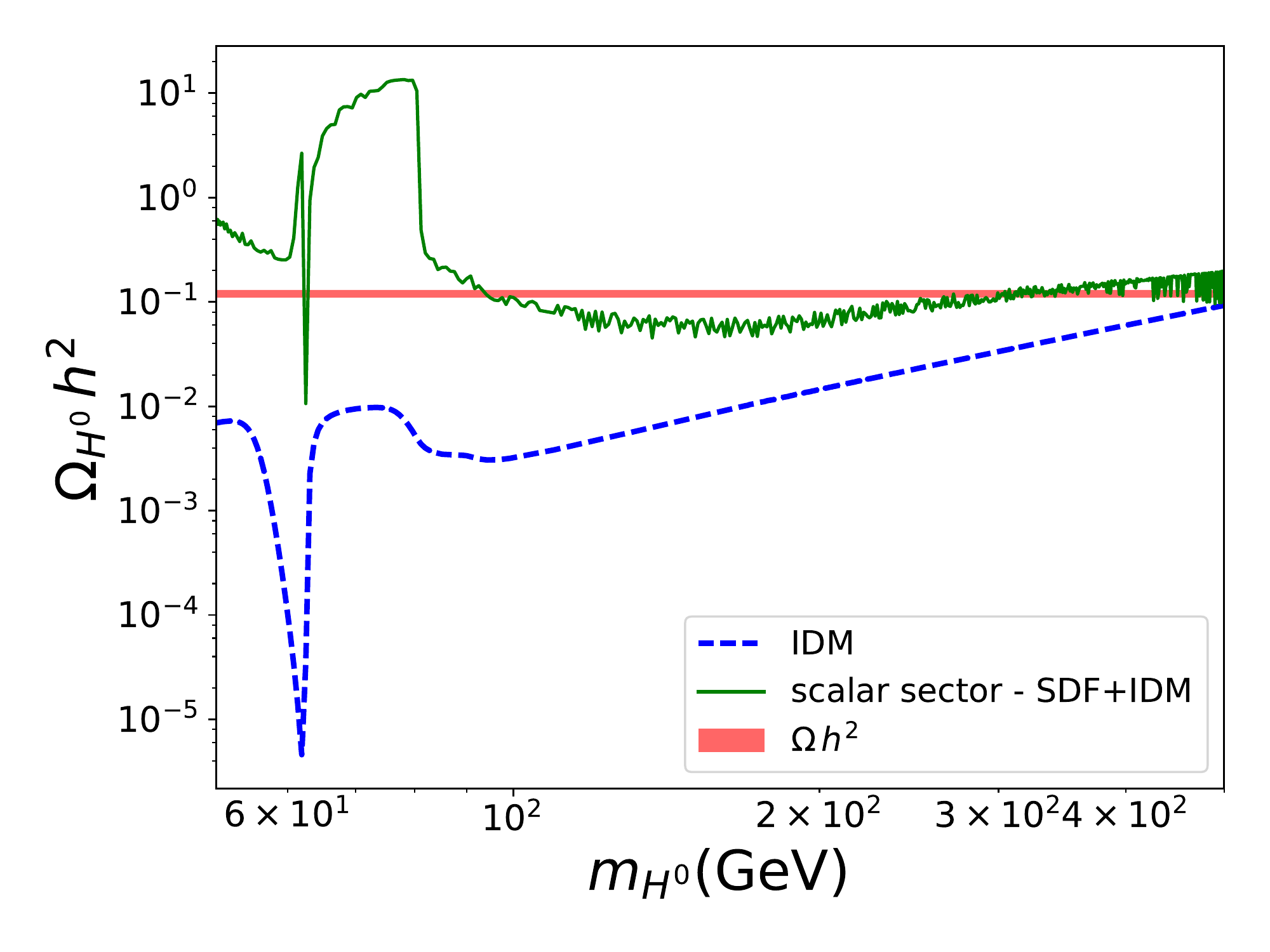}
\caption{Example of DM conversion in the SDFDM$+$IDM model. We choose the parameters $\lambda_L=0.045, y_2=10^{-2}, y_1=10^{-3}$ $m_{\chi_1}- m_{H^0}\, \lesssim\, 1.1 $ GeV and $m_{\Psi}>1$ TeV.}
{\label{fig:SDFDM-DM-conversion}}
\end{figure}
The complete model is given by the combination of the SDFDM and the IDM model. It will be dubbed as SDFDM$+$IDM for short. There are two DM candidates, the Majorana fermion $\chi_1^0$ of the SDFDM model and the scalar field $H^0$ of the IDM model
%~\footnote{Without loss of generality, we choose $m_{H^0}<m_{A^0}.$}. 
Now, with two DM particles, we need to take into account that in the early Universe, DM conversion could change the abundance for each specie as was suggested in Sec.~\ref{sec:two_DM}. 
In Fig.~\ref{fig:SDFDM-DM-conversion} we show an example in which the scalar abundance of $H^0$ is enhanced by the annihilation of the fermion field $\chi_1^0$. The blue dashed-line shows the typical behavior of the IDM model for some specific parameters. However, when we add the fermion field $\chi_1^0$, the DM abundance is enhanced to the green solid line. This behavior is obtained because we have over-abundance of fermion $\chi_1^0$ for the parameters that we fixed in Fig.~\ref{fig:SDFDM-DM-conversion}. Therefore, the process $\chi_1^0\bar{\chi}_1^0\to H^0H^0$ is opened as we described in Sec.~\ref{sec:two_DM} and enhance the relic abundance for the scalar particle $H^0$ in the early Universe.

\subsection{Numerical results}

In order to do a complete analysis of the SDFDM$+$IDM model, we carry out a scan of its parameter space as is shown in Table~\ref{table:SDFDM_parameter_space}.
We implemented the model in~\texttt{SARAH}~\cite{Staub:2008uz,Staub:2009bi,Staub:2010jh,Staub:2012pb,Staub:2013tta}, coupled to the \texttt{SPheno}~\cite{Porod:2003um,Porod:2011nf} routines. 
To obtain the DM relic density, we used~\texttt{MicrOMEGAs}~\cite{Belanger:2006is}, which takes into account all the possible channels contributing to the relic density, including processes such as coannihilations and resonances~\cite{Griest:1990kh}. We selected the models that can account for the total $\Omega h^2$ to  $3\sigma$ standard deviation
according to Planck satellite measurement~\cite{Aghanim:2018eyx}, as well as the constraints described in Sec.~\ref{sec:pheno_constraints}.
For those points, we computed the SI DM-nucleus scattering cross section, and checked it against the current experimental bounds of XENON1T~\cite{Aprile:2018dbl}, and prospect bounds for DARWIN~\cite{Aalbers:2016jon}, the most sensitive DD experiment planned.
\begin{table}
\centering
\begin{tabular}{|c|c|} 
\hline
Parameter & Range\\
\hline
$M_N$  & $10^{0}-5\times10^{3}$ (GeV) \\
$M_{\Psi}$  & $10^{2}-5\times10^{3}$ (GeV) \\
$y_{1,2}$ & $10^{-4}-10$ \\
\hline
\end{tabular}
\caption{Scan range of the parameters of the SDFDM model. The parameters of the IDM model are scanned as is shown in Table~\ref{table:IDM_scan}.}
\label{table:SDFDM_parameter_space}
\end{table}

\subsection{Relic density}

\begin{figure}
\includegraphics[scale=0.38]{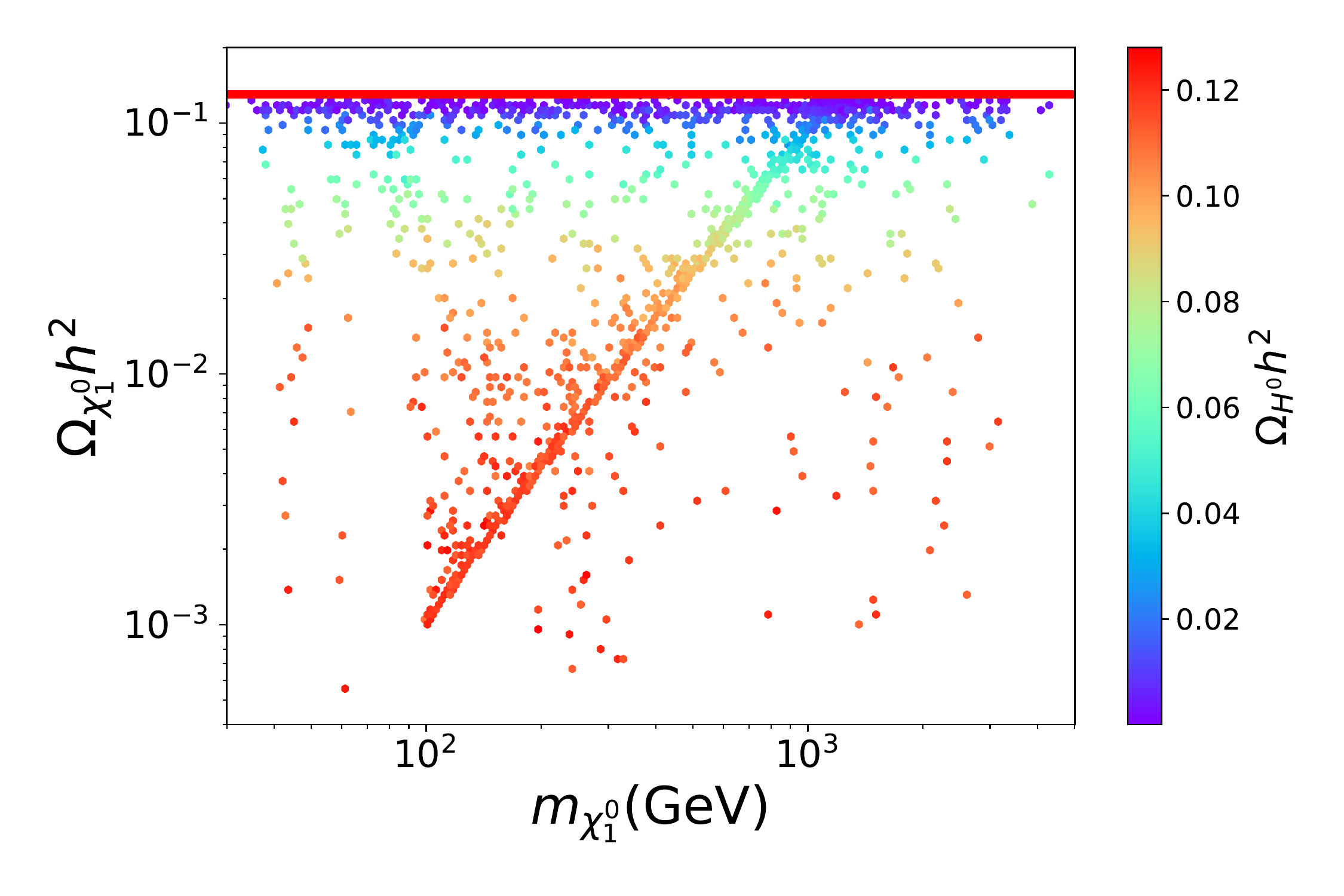}
\includegraphics[scale=0.38]{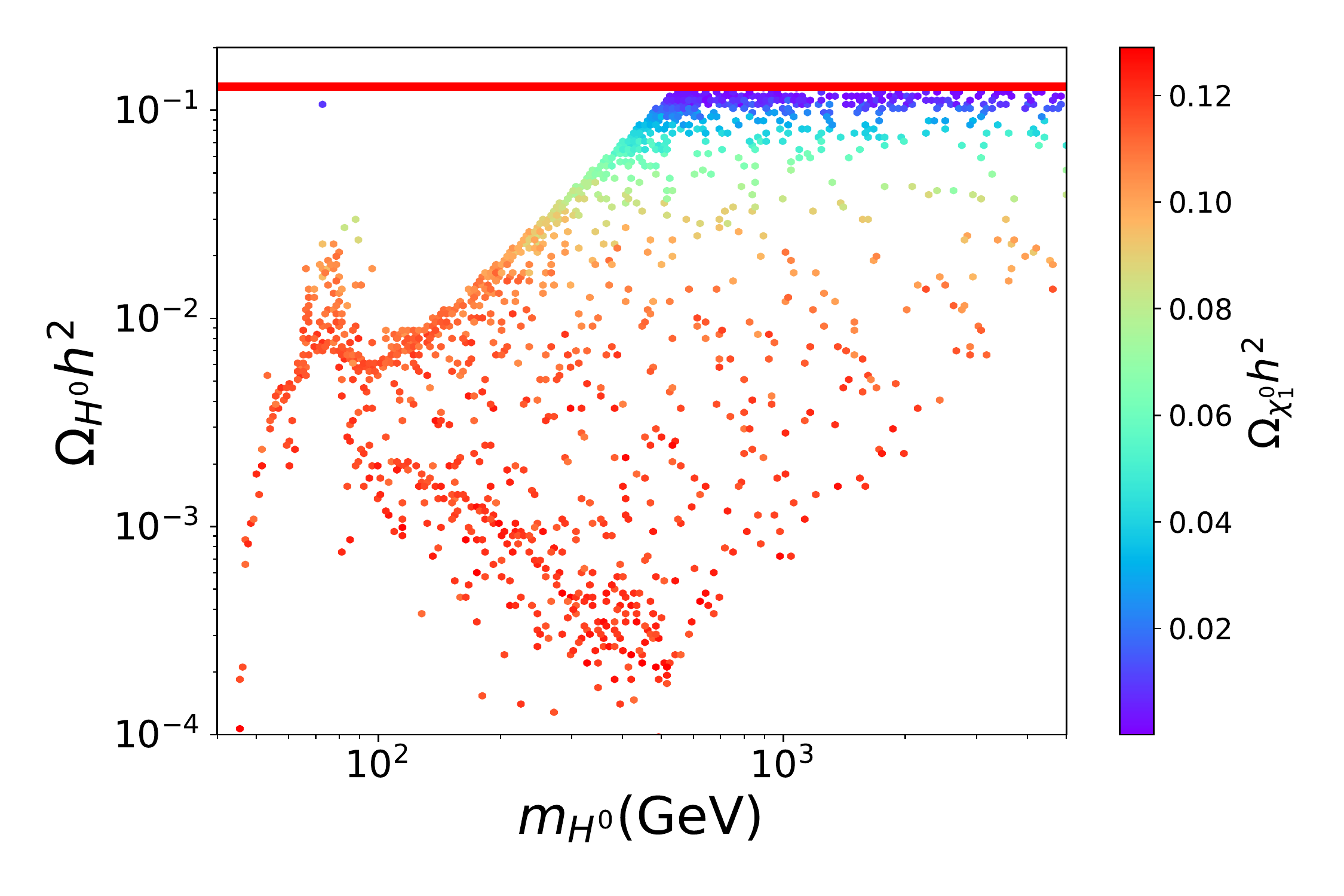}
\caption{Left: Fermion abundance. Right: Scalar abundance. All those models fulfill that $(\Omega_{H^0}+\Omega_{\chi_1^0})h^2=(0.1200\pm 0.0036)$ to $3\,\sigma$ according Sec.~\ref{sec:pheno_constraints} (see eq.~\eqref{eq:exper_relic_abundance}).}
{\label{fig:DM-abundance}}
\end{figure}

In the SDFDM$+$IDM model, DM conversion could alter the abundance of each species as was shown in Sec.~\ref{sec:SDFDM-conversion}. 
However, we checked that when we impose the experimental constraint on the relic abundance to $3\,\sigma$, this effect is not sizeable for this model and the DM conversion does not
play an important role. This is because $\sigma_{v}^{iijj}$ is smaller than $\sigma_{v}^{ii}$, and therefore, the relic abundance is obtained for each model with a negligible  communication in the early Universe.
On the left side of Fig.~\ref{fig:DM-abundance} we show the DM abundance for the fermion field $\chi_1^0$. 
The blue points show that the SDFDM model itself could account for the observed DM abundance without the contribution of the scalar field $H^0$.
Also, on the right side of Fig.~\ref{fig:DM-abundance} we show the DM abundance for the scalar field $H^0$. We note that it is always below the experimental value for $m_{H^0}\lesssim 550$ GeV, except for points near to the resonance with the SM Higgs field, which is the known behavior of the IDM model. 
For $m_{H^0} \gtrsim 550$ GeV, the IDM model can explain the total value of the relic abundance (blue points). However, for $m_{H^0}\lesssim 550$ GeV the presence of the fermion component $\chi_1^0$ is necessary in order to obtain the experimental value for $\Omega\,h^2$ (red points).

\subsection{Direct detection}

At tree-level, the SDFDM$+$IDM model has nucleon recoil signals. 
The fermion $\chi_1^0$ and the scalar field $H^0$ interact with nucleons through the Higgs field of the SM and also through the $Z$ gauge boson portal. 
For the fermion DM component, the SI interaction through the scalar portal gives a cross section
\begin{equation}
\label{eq:SI-SDFDM}
\sigma^{SI}_{\chi_1^0} \approx 2\dfrac{m_r^2}{\pi} \left(\dfrac{c_{\chi_1^0\chi_1^0 h}}{ v m_h^2}\right)^2f_N^2 m_N^2\,,
\end{equation}
where $c_{\chi_1^0\chi_1^0 h} = \sqrt{2} \mathbf{O}_{11}(y_1 \mathbf{O}_{12}-y_2 \mathbf{O}_{13})$ is the coupling between the DM and the Higgs scalar field, $m_r=m_N m_{\chi_1^0}/(m_N +m_{\chi_1^0})$ is the reduced mass, and $f_N \approx 0.3$ is the form factor for the scalar interaction~\cite{Alarcon:2011zs,Alarcon:2012nr}.
Also, for the scalar DM component, the SI cross section is given by eq.~\eqref{eq:IDM_SI}.
 
Before presenting our results, it is important to point out that for the case of multicomponent DM, the constraints coming from DD do not apply directly. This happens because DM-nucleon recoil rates are dependent on the local density of the DM candidate. 
\begin{figure}
\includegraphics[scale=0.38]{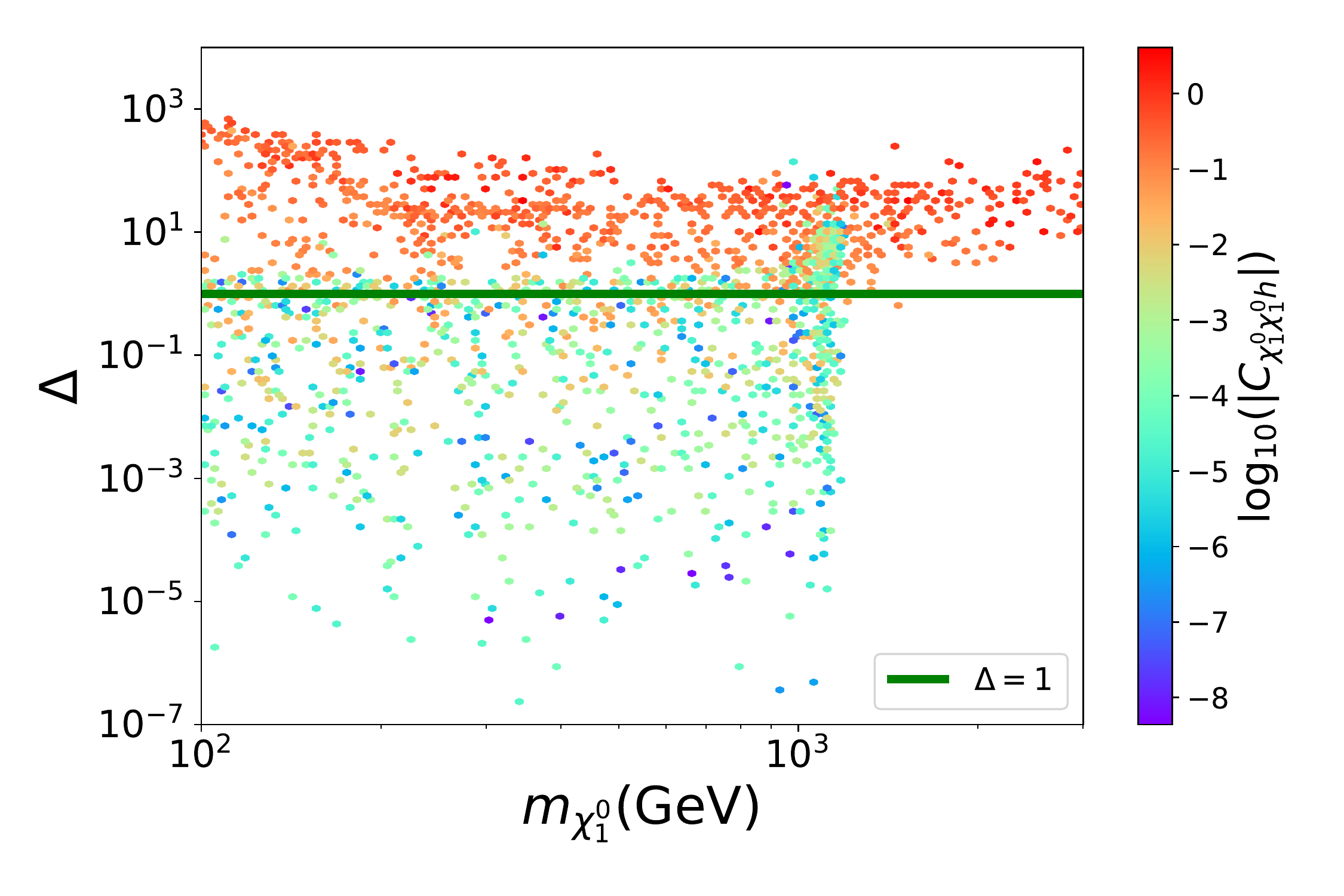}
\includegraphics[scale=0.38]{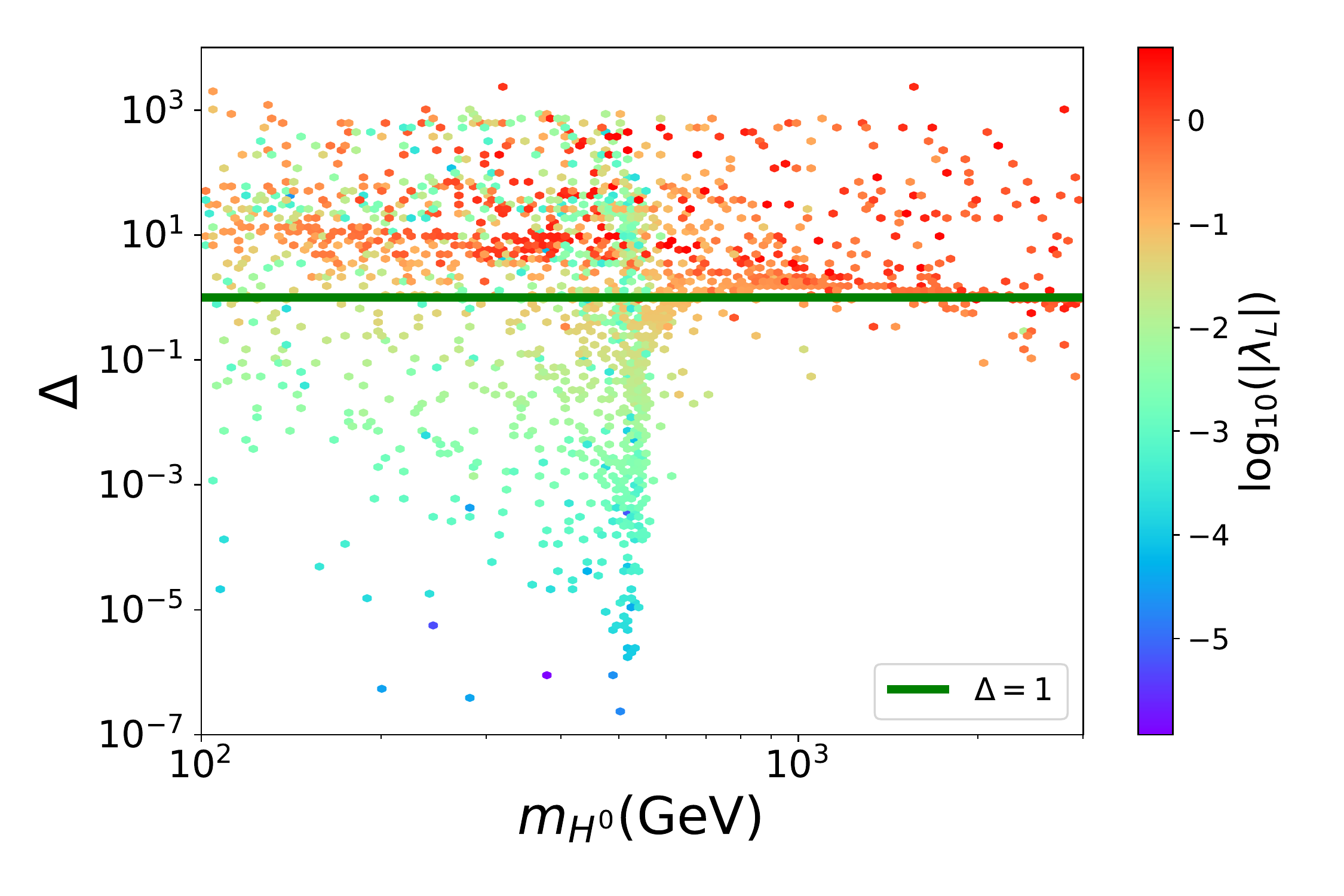}
\caption{Value for $\Delta$  as a function of the dark matter mass for each specie, $m_i$, $ i \in \lbrace \chi_1^{0}, H^0 \rbrace $. 
The green line ($\Delta = 1 $) represents the current upper limit on the $\Delta$ due to XENON$1\rm{T}$~\cite{Aprile:2018dbl} restrictions. Points above the green light are ruled out.}
{\label{fig:SI-SDDMF}}
\end{figure}
In order to account for this, the cross section for each DM candidate must be re-scaled by the $\Omega_i/\Omega$ factor, where $\Omega_i h^2$ is the relic abundance for the $\chi_1^0$ or $H^0$ field and $\Omega h^2$ is the experimental value described in Sec.~\ref{sec:pheno_constraints}. 
For multicomponent DM, the scattering cross section for each DM candidate must be rescaled and limits from DD experiments adjusted, thus the restrictions can be placed instead on the parameter $\Delta$~\cite{Cao:2007fy}:
\begin{equation}
   \Delta = \dfrac{\sigma^{SI}_{H^0}}{\sigma^{SI}_{X_e}(M_{H^0})}\left(\dfrac{\Omega_{H_0}}{\Omega}\right) + 
    \left(\dfrac{\sigma^{SI}_{\chi_1^0}}{\sigma^{SI}_{X_e}(m_{\chi_1^0})}
    +
    \dfrac{\sigma^{SD}_{\chi_1^0}}{\sigma^{SD}_{X_e}(m_{\chi_1^0})}\right)\left(\dfrac{\Omega_{\chi_1^0}}{\Omega}\right) <1\,,
    \label{eq:delta}
\end{equation}
where $\sigma^{SD}_{\chi_1^0}$ is the SD cross section.
In  Fig.~\ref{fig:SI-SDDMF} it is shown $\Delta$ as a function of the mass of each DM species.
Although points with large $\lambda_L$ and large $c_{\chi_1^0\chi_1^0 h}$ generate a huge SI cross section and they could exceed the XENON1T limit, they can not be excluded because they could have a low contribution to the relic density of the DM. The interplay between the relic density, the SI and SD cross section needs to be taken into account as it is shown by the $\Delta$ parameter.

\subsection{Indirect detection}

\begin{figure}
\includegraphics[scale=0.38]{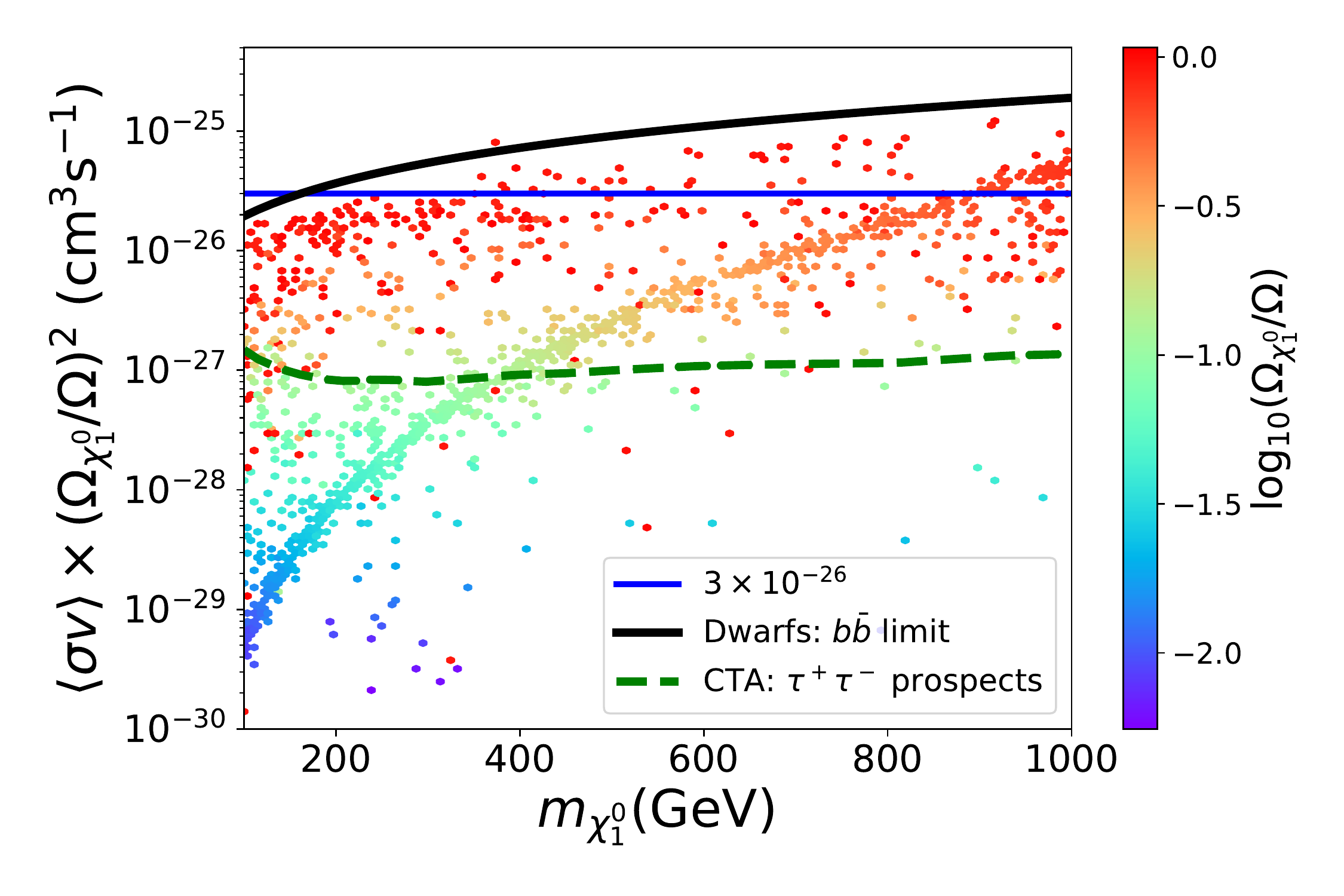}
\includegraphics[scale=0.38]{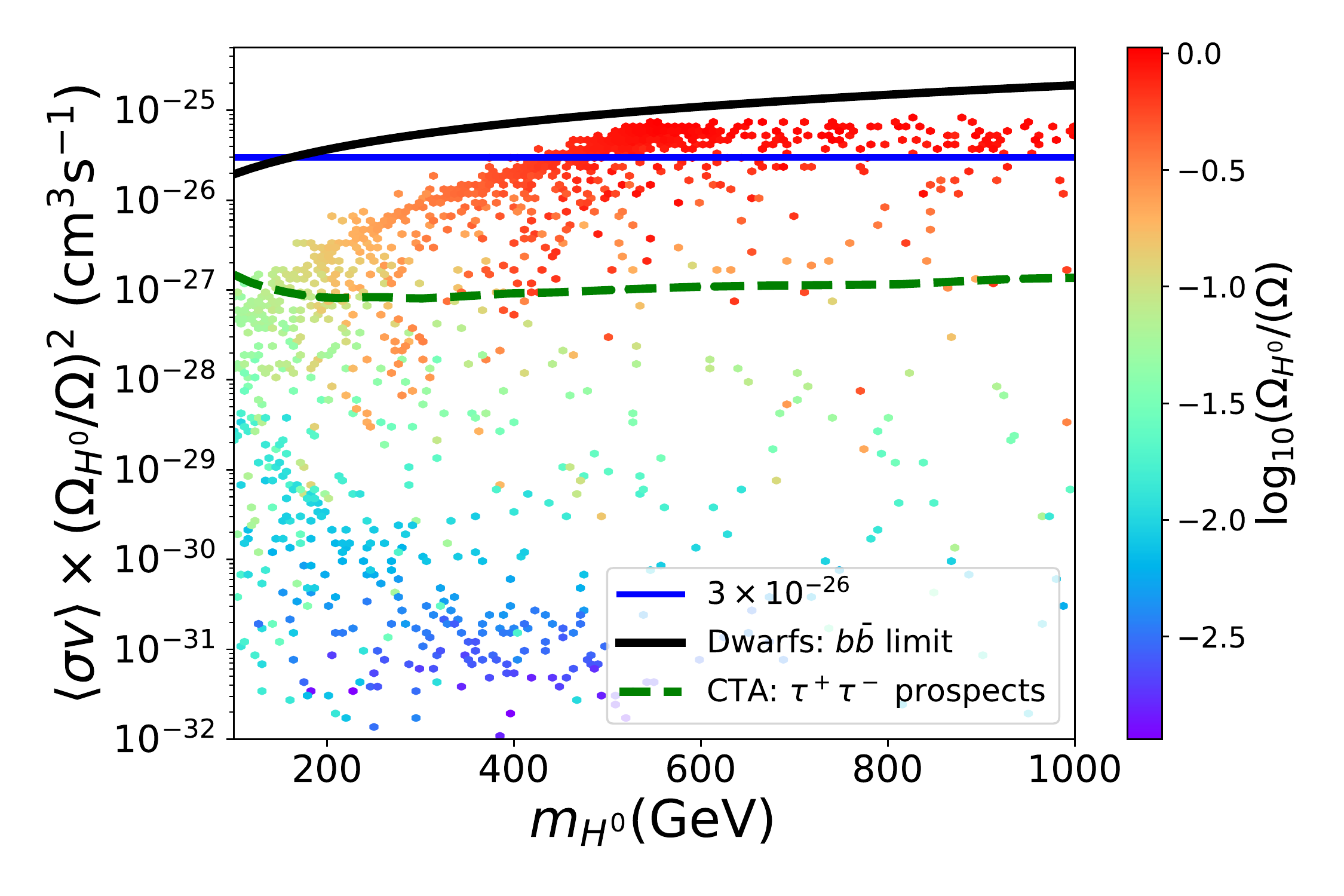}
\caption{Thermally averaged annihilation cross section today. Left: We show the re-scaling with $(\Omega_{\chi_1^0}/\Omega)^2$ for the fermion DM component. Right: the same for the scalar DM component. We also show the typical thermal value $\langle \sigma v\rangle \sim 3\times 10^{-26} \text{cm}^3 \text{s}^{-1}$ in the early Universe, the experimental limit for DM annihilation into $b\bar{b}$ in dwarf galaxies~\cite{Ackermann:2015zua} and CTA prospects for DM annihilation into $\tau^+\tau^-$ channel~\cite{Funk:2013gxa,Wood:2013taa}.}
{\label{fig:SDFDMsv}}
\end{figure}

In Fig.~\ref{fig:SDFDMsv} we show the thermally averaged annihilation cross section $\langle \sigma v\rangle$ for the SDFDM$+$IDM model. Similar to the case of DD, for this observable we must rescale the $\langle \sigma v\rangle$ by the factor $(\Omega_{\chi_1^0}/\Omega)^2$ for the fermion DM particle and $(\Omega_{H^0}/\Omega)^2$ for scalar DM component.  Our results show that the models are always under the current Fermi-LAT limits even in the better case for a large branching ratio of the annihilation channels
$\chi_1^0\bar{\chi}_1^0 \to b\bar{b}$ or $H^0\,H^0 \to b\bar{b}$, which leads to DM annihilation into $b\bar{b}$ signal from dwarf galaxies (dSphs)~\cite{Ackermann:2015zua}. 
In color, we also show the behavior of the relic density for both figures. We realize that a sizeable amount of DM demands a high $\langle \sigma v\rangle$. 
Also, we find that the $(\Omega_{\chi_1^0}/\Omega)^2$ and $(\Omega_{H^0}/\Omega)^2$ factors controls the thermal velocity annihilation cross section. Therefore, It demands low gamma-ray fluxes, all under the the current Fermi-LAT limits for DM annihilation in dwarf galaxies~\cite{Ackermann:2015zua}.
We also find that a region of the parameter space could be probed by next generation of experiments such as CTA (green dashed curve) for DM annihilation into $\tau^+\tau^-$ channel~\cite{Funk:2013gxa,Wood:2013taa}.

\subsection{Collider phenomenology}

The LHC has reached staggering energies and number of collisions. Thus, it is possible, in principle, to explore the model with the ATLAS and CMS experiments. The restrictions that could rise from the energy frontier are dependent on the allowed topologies which in turn depend on the mass splittings of the dark sector.

In the case of the IDM the collider constraints have been explored extensively in the literature. For instance, in \cite{Datta:2016nfz} the discovery prospects on multilepton channels and 3000~$\rm{fb}^{-1}$ luminosity was studied, while in \cite{Poulose:2016lvz} the two jets plus missing transverse energy signal was explored. More recently, \cite{Belyaev:2016lok} studied IDM signatures such as Mono-jet, Mono-Z and Mono-Higgs production and vector boson fusion.  Since there are many dedicated works for the IDM exploring its rich collider phenomenology, we will focus on the collider prospects of the fermion content.

\begin{figure}
 \centering
 \includegraphics[scale=0.38]{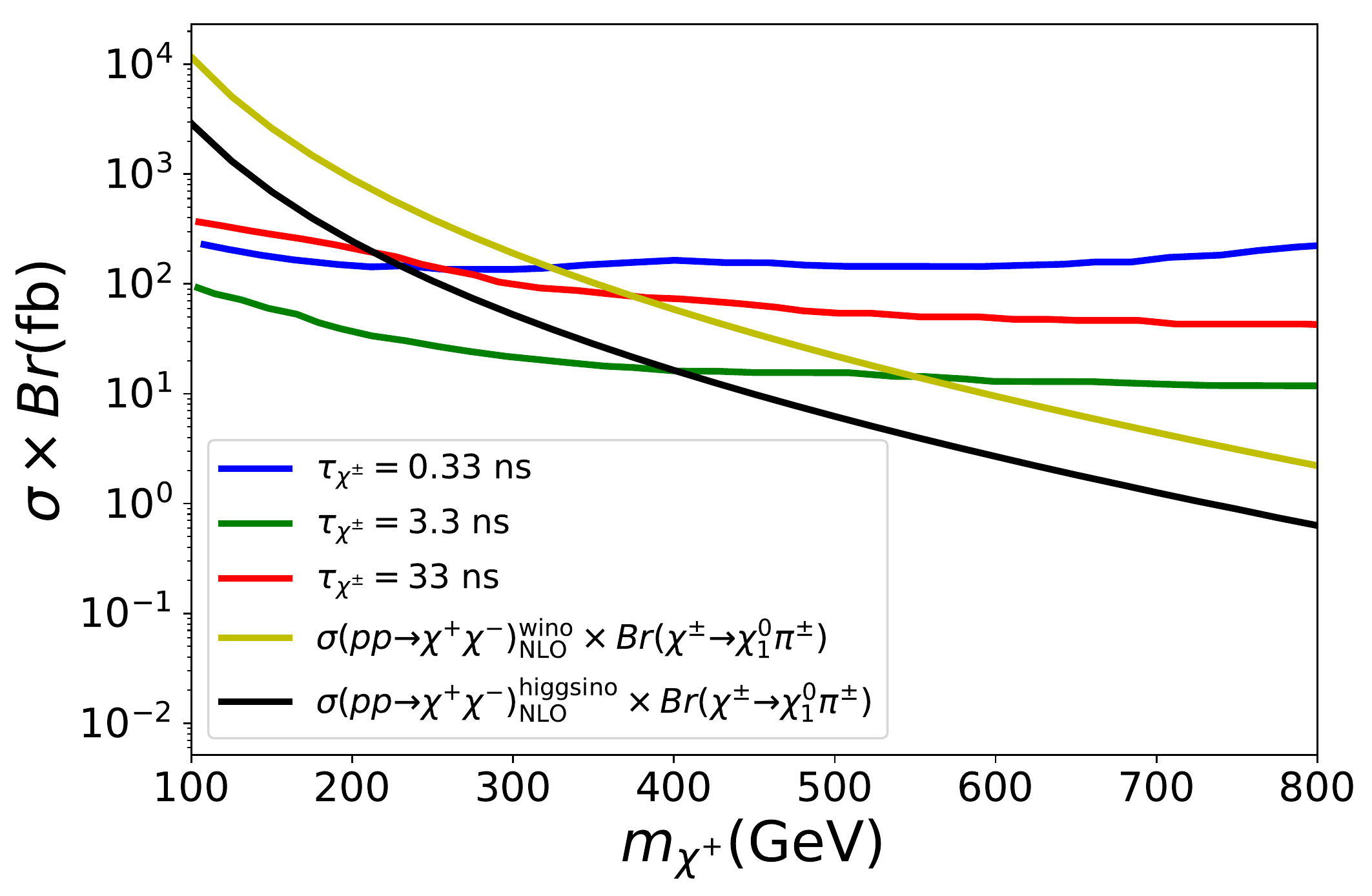}
 \caption{ The solid blue, red and green lines
   are the observed  $95\%$ CL upper limits
   of the product $\sigma(pp \to \chi^{+} \chi^{-} ) \times Br(\chi^{\pm} \to  \chi_{1}^{0} \pi^{\pm})$
   as a function of $m_{\chi^{+}}$ for wino like fermions with lifetimes of  $0.33$~ns, $3.3$~ns,
   and $33$~ns respectively~\cite{Sirunyan:2018ldc}.
   The solid black   represent the theoretical
   expression for the product
   $\sigma(pp \to \chi^{+} \chi^{-} )_{\rm{NLO}} \times Br(\chi^{\pm} \to  \chi_{1}^{0} \pi^{\pm})$
   as a function of $m_{\chi^{+}}$.
}
{\label{fig:stf_width_cms}}
\end{figure}

For SDFDM model, after imposing the aforementioned constraints, we find that the mass splitting between the lightest charged fermion and the fermionic DM is very small. In fact, most points are in the mass splittings of $m_{\pi^{\pm}}  <(m_{\chi^{+}} - m_{\chi^{0}_1}) < 0.5~\rm{GeV}$, where $m_{\pi^{\pm}}=139.6~\rm{MeV}$ is the charged Pion mass. In this case, the most predominant decay mode of charged fermion
is $\chi^{\pm} \to \pi^{\pm} \chi_1^{0} $, with $Br(\chi^{\pm} \to \pi^{\pm} \chi_1^{0} ) \geq 0.97 $,
however, the charged fermion $\chi^{\pm}$ has a small width decay, allowing it to 
travel inside the detector before decay~\cite{Sirunyan:2018ldc}. In  the CMS analysis~\cite{Sirunyan:2018ldc}, a
search of long-lived charginos in a supersymmetry model
is carried out, 
using disappearing track signatures
and exclude
charginos with lifetimes from $0.5$~ns to $60$~ns for chargino
masses of $505$~GeV.
This analysis has the potential to put constraints
in a small region of the parameter space of the model. In the Fig.~\ref{fig:stf_width_cms}
is shown the 2$\sigma$ upper experimental limits on production cross section
times branching ratio  for wino-like chargino pairs for three different lifetimes. The solid
black line represents the theoretical cross section for the model prediction
in the limit when the charged fermion is mostly doublet.
Charged fermions with masses of $210$ GeV, $220$ GeV and $400$ GeV are excluded
for lifetimes of $33$~ns, $0.33$~ns  and $0.3$~ns  respectively.
%
%%%%%%%%%% BP %%%%%%%%%%%%%%
\subsection{Benchmark points}
\label{BP_DTF}
In this section, we include two benchmark points of the model that satisfy the constrains metioned in the previous sections:

\begin{center}
	\begin{tabular}{|l |l| l| l|}
		
		\hline
		
		BP &	Scalar Parameters    & Fermion Parameters & Observables \\
		\hline
		
		BP1&	$\mu_2= $ 202.12 GeV      & $M_{N}=$ 2277.5 GeV   & $\Omega_{\chi_1^0} h^2=0.103$      \\
		& $\lambda_3=$ -3.63$\times 10^{-4}$        & $M_{\Psi}=$ 1058.9 GeV & $\Omega_{H^0} h^2=0.015$      \\
		&	$\lambda_4=$ -7.53$\times 10^{-2}$       & $y_1=$1.62$ \times 10^{-4}$     & $\Delta=$0.56     \\
		& $\lambda_5=$ -1.31$\times 10^{-3}$ & $y_2=$1.19 $\times 10^{-2 	}$ & \\
		\hline
		\hline
		BP2& $\mu_2= $ 560.64 GeV   & $M_{N}=$ 182.3 GeV     & $\Omega_{\chi_1^0} h^2=$ $1.69 \times 10^{-3}$     \\
		& $\lambda_3=$ 8.07 $ \times 10^{-2}$        & $M_{\Psi}=$ 124.1 GeV & $\Omega_{H^0} h^2=$  0.119    \\
		&	$\lambda_4=$ -3.83 $\times 10^{-2}$       & $y_1=$ 9.73 $\times 10^{-2}$     & $\Delta=$  $5.23\times 10^{-2}$   \\
		& $\lambda_5=$ -5.12 $\times 10^{-3}$ & $y_2=$ 7.33 $ \times 10^{-2}$ & \\
		\hline	
	\end{tabular}
\end{center}

\section{Doublet-Triplet Fermion Dark Matter Model}
\label{sec:fermionsector}
%%%%%%%%%% MODEL %%%%%%%%%%%%%%
In the doublet-triplet model (DTF), the fermionic sector of the SM is enlarged by adding an  $SU(2)_L$ vector-like doublet and a Majorana triplet, both being odd under the $Z_2'$ symmetry. In order to express the most general renormalizable Lagrangian, the masses, and interactions, we will closely follow the notation of ~\cite{Freitas:2015hsa}, thus, the new fields are:
\begin{align}\label{eq:fermioncontent}
\psi_{L}=\left( \begin{array}{ccc}
\psi^0_{L}  \\
\psi^-_{L} \end{array} \right),\hspace{1cm} 
\psi_{R}=\left( \begin{array}{ccc}
\psi^0_{R}  \\
\psi^-_{R} \end{array} \right),\hspace{1cm}
\Sigma_L\equiv\sqrt{2}\Sigma^i_L\tau^i=\left( \begin{array}{ccc}
 \Sigma_L^0/\sqrt{2} &  \Sigma_L^+\\
 \Sigma_L^{-} & -\Sigma_L^0/\sqrt{2} \end{array} \right),
\end{align}
where $\tau^i=\sigma^i/2$, $\Sigma^{\pm}_L\equiv(\Sigma_L^1\mp i\Sigma_L^2)/\sqrt{2}$ and $\Sigma_L^3=\Sigma_L^0$. 
The part of the Lagrangian containing the kinetic and mass terms for the new fields reads:
\begin{align}
\mathcal{L}_F={\rm Tr}[\bar{\Sigma}_L i\gamma^\mu D'_\mu \Sigma_L]-\frac{1}{2}{\rm Tr}(\bar{\Sigma}_L^cM_\Sigma\Sigma_L+\mbox{h.c.})+\bar{\psi} i\gamma^\mu D_\mu\psi-M_\psi(\bar{\psi}_R\psi_L+\mbox{h.c.}).
\end{align}
These terms are in agreement with ~\cite{Freitas:2015hsa,Biggio:2011ja}.
On the other hand, the new fermions can not mix with SM leptons due to the $Z_2'$ symmetry. Thus, the most general Yukawa Lagrangian only involves interactions with the Higgs boson:
\begin{align}
  \label{eq:YukawaLagrangian}
  \mathcal{L}_Y&=  -y_1H^\dagger\overline{\Sigma_L^c}\epsilon \psi_R^c   +  y_2 \overline{\psi_L^c} \epsilon \Sigma_L H +  {\rm h.c.}\\
&=-\frac{h+v}{2} \left[y_1\left(\overline{\Sigma_L^{0c}}\psi_R^{0c}+\sqrt{2}\overline{\Sigma_L^{-c}}\psi_R^{-c}\right)+y_2\left(\overline{\psi_L^{0c}}\Sigma_L^0+\sqrt{2}\overline{\psi_L^{-c}}\Sigma_L^+\right)+  {\rm h.c.}\right]\,,
\end{align}
where  $y_{i}$ are Yukawa couplings controlling the new interactions and $H=(0,\,(h+v)/\sqrt{2})^T$, $h$ being the SM Higgs boson and $v=246$ GeV is the VEV.
Once the electroweak symmetry is spontaneously broken the $y_{i}$ terms generate a mixture in the neutral and charged sectors leading to a mass matrix in the basis $\Xi^0=(\Sigma_L^0, \psi^0_L, \psi^{0c}_R)^T$ and to a charged fermion mass matrix in the basis $\Xi^-_R=(\Sigma^{+c}_L, \psi_R^{-})^T$ and $\Xi^-_L=(\Sigma^{-}_L,\psi^-_L)^T$ given by:
\begin{align}
\label{eq:Mchi2}
  \mathbf{M}_{\Xi^0}=\begin{pmatrix}
 M_\Sigma                 &\frac{1}{\sqrt{2}}yv\cos\beta& \frac{1}{\sqrt{2}}yv\sin\beta\\
\frac{1}{\sqrt{2}}yv\cos\beta &  0                  & M_\psi\\
\frac{1}{\sqrt{2}}yv\sin\beta&  M_\psi                &  0  \\
\end{pmatrix},\hspace{1cm}
  \mathbf{M}_{\Xi^\pm}=\begin{pmatrix}
 M_\Sigma         &   yv\cos\beta \\
 yv\sin\beta & M_\psi \\
\end{pmatrix}.
\end{align}
Here we have defined $y=\sqrt{(y_1^2 + y_2^2)/2}$ and $\tan\beta=y_2/y_1$. 
Similar to the case of the SDFDM, in this model, the fermionic neutral mass eigenstates are obtained via $\mathbf{O}^{\operatorname{T}}\mathbf{M_\Xi^0} \mathbf{O}=\mathbf{M}^\chi_\text{diag}$ while the charged ones are obtained through $\mathbf{U_{L}}^{\operatorname{T}}\mathbf{M_\Xi^{\pm}} \mathbf{U_{R}}=\mathbf{M}^{\chi^{\pm}}_\text{diag}$. As a result, the mass eigenstates includes three neutral Majorana states, namely $\chi_1^0$, $\chi_2^0$ and $\chi_3^0$,  and two charged fermion particles $\chi_1^\pm$ and $\chi_2^\pm$. Due to the $Z_2'$ symmetry the lightest neutral fermion is stable and therefore the fermionic dark matter candidate. In this notation, we assume the mass ordering $|m_{\chi_1^{\pm}}| < |m_{\chi_2^{\pm}}|$ and $|m_{\chi_1^0}| < |m_{\chi_2^0}| < |m_{\chi_3^0}|$ thus the fermionic DM field is $\chi_1^0$.
Though the DTF model presents an interesting phenomenology, its DM candidate is underabundant on most of the parameter space. For this reason we consider also de IDM, such that the model has two DM candidates.

\subsection{DM conversion}
\label{DMconversion_DT}

\begin{figure}[h]
\centering
\includegraphics[scale=0.38]{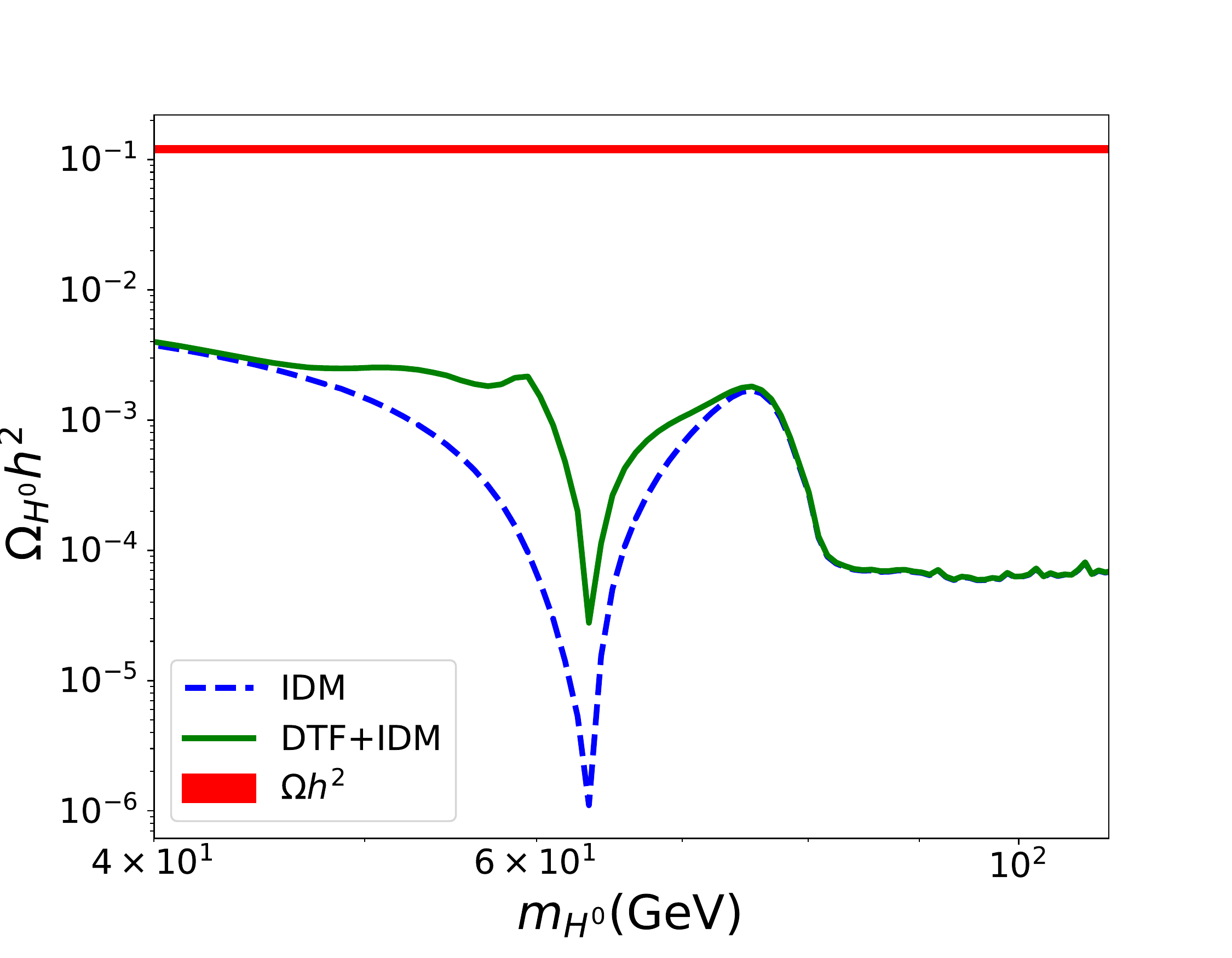}
\caption{Relic density dependence on the $m_{H^0}$ mass for the IDM (blue dashed), and the scalar sector of the model (green solid). The difference between the two curves shows that in the multicomponent scenario, DM conversion is playing a role in the relic abundance. The red band represents the observed relic abundance.}
\label{fig:DT_DM_conversion}
\end{figure}

In this model, in order to keep the $Z_2$ and $Z_2'$ symmetries exact, there are a few ways the two sectors may communicate. Nevertheless, it is possible to have two scalar (fermionic) DM particles converting into two fermionic (scalar) DM particles, for instance, through an s-channel annihilation. In order to understand the impact that this DM conversion has on the total relic density we studied the annihilation through the Higgs portal for specific parameters of the model. In this case we set $y_1=1.0$, $y_2=1.5$ and $\lambda_L=0.7$, we also vary the mass of the scalar DM particle $H_0$ and keep the fermion DM mass such that $2 \ \mathrm{GeV}<|m_{\chi_1^0}|-m_{H^0}<7$ GeV. The results are shown in Fig. \ref{fig:DT_DM_conversion} where a clear difference in the relic abundance is found between just the IDM (blue dashed curve) and the scalar sector of the full model (green solid), this is due to the DM conversion between the two sectors. It is worth noting that the curves stop differing at masses that are larger than the weak gauge boson masses. This is because for such masses, the annihilation through the t and u channel exchange of weak gauge bosons dominates and the impact of the Higgs portal is suppressed. 

%%%%%%%%%%%% FIGURE%%%%%%%%%%%%%%%%%
%%%%%%%%%%%% FIGURE%%%%%%%%%%%%%%%%%

\subsection{Numerical results}

The DTF+IDM presents an interesting phenomenology, thus, in order to study it, we performed a scan of the parameter space as is shown in Table~\ref{table:DTF_parameter_space}. The model was impleted in \texttt{SARAH} ~\cite{Staub:2013tta} and connected to \texttt{SPheno}. The output was then exported to \texttt{MicrOMEGAs} ~\cite{Belanger:2018mqt} in order to obtain the two-component relic density, the SI cross section, and the thermally averaged annihilation cross section of both candidates. For collider constraints we exported the model to the Monte Carlo generator \texttt{MadGRAPH (v5.2.5.5)}. 
The new fields within the model have the potential of affecting precision observables such as the $S$ and $T$, parameters and $R_{\gamma \gamma}$. To this end, in the following sections we only present results that satisfy all the constraints presented in Sec. \ref{sec:pheno_constraints}, except for the left side of Fig. ~\ref{fig:DTFDM_SandFrelic} where the phenomenology that leads to the correct relic abundance is interesting enough to be presented.
\begin{table}[h]
\centering
\begin{tabular}{|c|c|} 
\hline
Parameter & Range\\
\hline
$M_{\psi}$  & $10^{0}-10^{3}$ (GeV) \\
$M_{\Sigma}$  & -$(10^{0}-10^{3})$ (GeV) \\
$y_{1,2}$ & $10^{-4}-3$ \\
\hline
\end{tabular}
\caption{ Range of the parameter scanned in the DTFDM+IDM model.}
\label{table:DTF_parameter_space}
\end{table}

%%%%%%%%%% RELIC DENSITY %%%%%%%%%%%%%%
\subsection{Relic density}
\label{relic_DTDM}
\begin{figure}[h]
\centering
\includegraphics[scale=0.38]{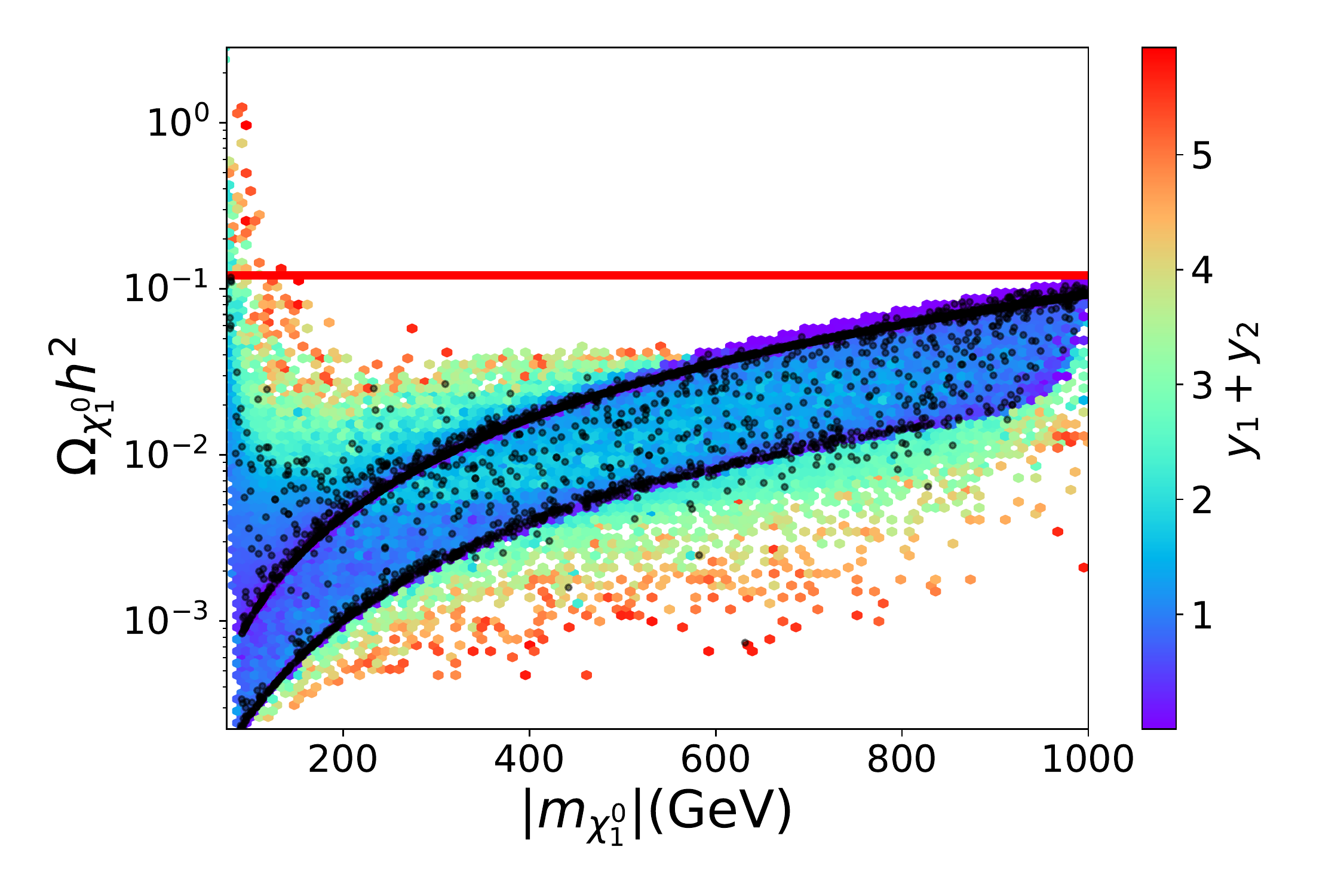}
\includegraphics[scale=0.38]{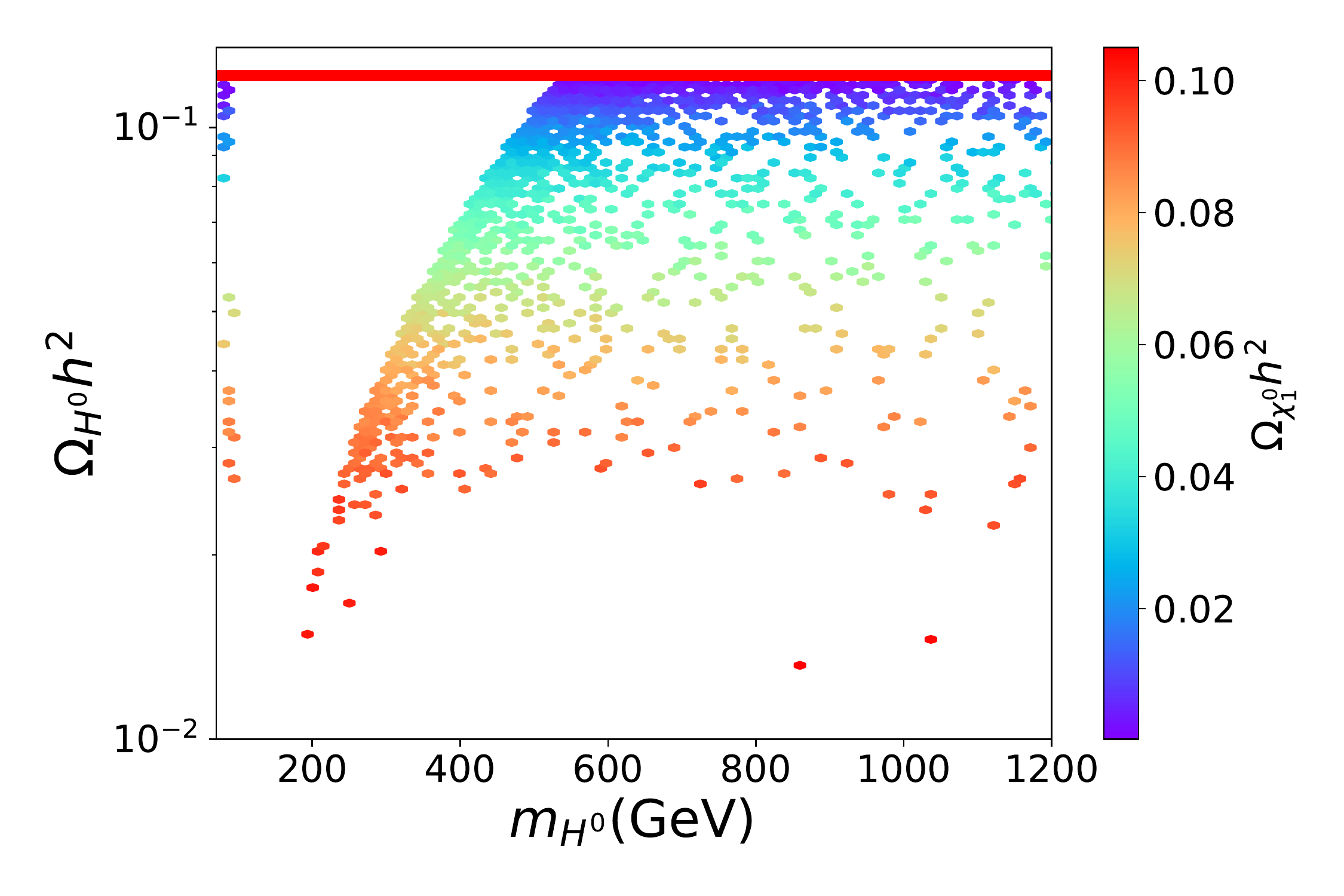}
\caption{Left side: Fermion  relic abundance vs. the fermion DM mass with the color gradient representing $y_1 + y_2$. Right side: Scalar relic abundance vs. the $m_{H^0}$ with the color gradient representing the fermion relic abundance. In this panel, all points fulfill the observed  $\Omega h^2$ at $3\,\sigma$.}
\label{fig:DTFDM_SandFrelic}
\end{figure}

In this model, due to the interplay of the fermionic DM and scalar DM sector, it is possible to saturate the relic abundance in most of the parameter space. 
The left side of Fig. \ref{fig:DTFDM_SandFrelic}, shows the fermionic relic abundance, $\Omega_{\chi_1^0}  h^2$ resulting from the scan versus the mass of the fermionic DM candidate, while the color gradient represents $y_1 + y_2$. The narrow red, horizontal band shows the allowed values of the relic density according to  \cite{Aghanim:2018eyx} with at most a 3$\sigma$ deviation from the central value. There are a few features of the plot that are worth considering. The black points represent those models that together with the scalar DM saturate the relic abundance. Most black points lie in two bands and those bands correspond to $y_1 + y_2 \sim 0$. Now, what happens at those small Yukawa values is that annihilation through the Higgs boson is suppressed which helps enhance the relic abundance. Moreover, the mass matrix diagonalization leads to nearly degenerate spectra thus, coannihilations play an important role. In fact, for the top band, there are more fermionic degenerate states, but due to the effective degrees of freedom, the annihilation cross section is less than that of the lower band. On the other hand, only for $|m_{\chi_{1}^{0}}| \sim  1.1$~TeV  it is possible for the fermion candidate to completely saturate the relic abundance, this is due to the high $SU(2)_L$ representation of the multiplets.

The right side of Fig. \ref{fig:DTFDM_SandFrelic} shows only points that satisfy the relic abundance in the $\Omega_{H^0} h^2-m_{H^0}$ plane while the color gradient represents the fermionic relic abundance. For $m_{H^0}$ between 100 GeV and 200 GeV the interplay of the two candidates does not saturate the correct abundance. Second, in the region $200$ GeV $<M_{H^0} < 500$ GeV, there is a very clear relation between the scalar DM mass and the fermion DM mass for values of $\Omega h^2$ near or at the observed value. This just shows that the suppressed abundance of one candidate must be overcome by the other candidate. However, in the region $M_{H^0} \gtrsim 550$ GeV it is possible to saturate the relic abundance just with the scalar sector of the model.

%%%%%%%%%%%% FIGURE%%%%%%%%%%%%%%%%%
%%%%%%%%%%%% FIGURE%%%%%%%%%%%%%%%%%

%%%%%%%%%% DIRECT DETECTION %%%%%%%%%%%%%%
\subsection{Direct detection}
\label{DD-DTF}
DD experiments are an interesting way to probe dark matter models, in fact in the case of WIMP dark matter, those experiments usually present some of the most stringent constraints. For the DTF model, the scattering of fermionic DM with nuclei occurs through Higgs exchange, and its approximate SI cross section is given by Eq.~\ref{eq:SI-SDFDM} where $c_{\chi_1^0 \chi_1^0 h}=   \mathbf{O}_{11}(y_1 \mathbf{O}_{12}-y_2 \mathbf{O}_{13})$.  The same happens to the IDM, thus, strong constraints could be expected. Moreover, the fermionic DM candidate also presents SD interactions mediated by the $Z$ boson. Nevertheless, for multicomponent DM, the scattering cross section for each DM candidate must be rescaled and limits from DD experiments adjusted, thus the restrictions can be placed instead on the parameter $\Delta$ presented in Eq. \ref{eq:delta}.

\begin{figure}[h]
	\includegraphics[scale=0.38]{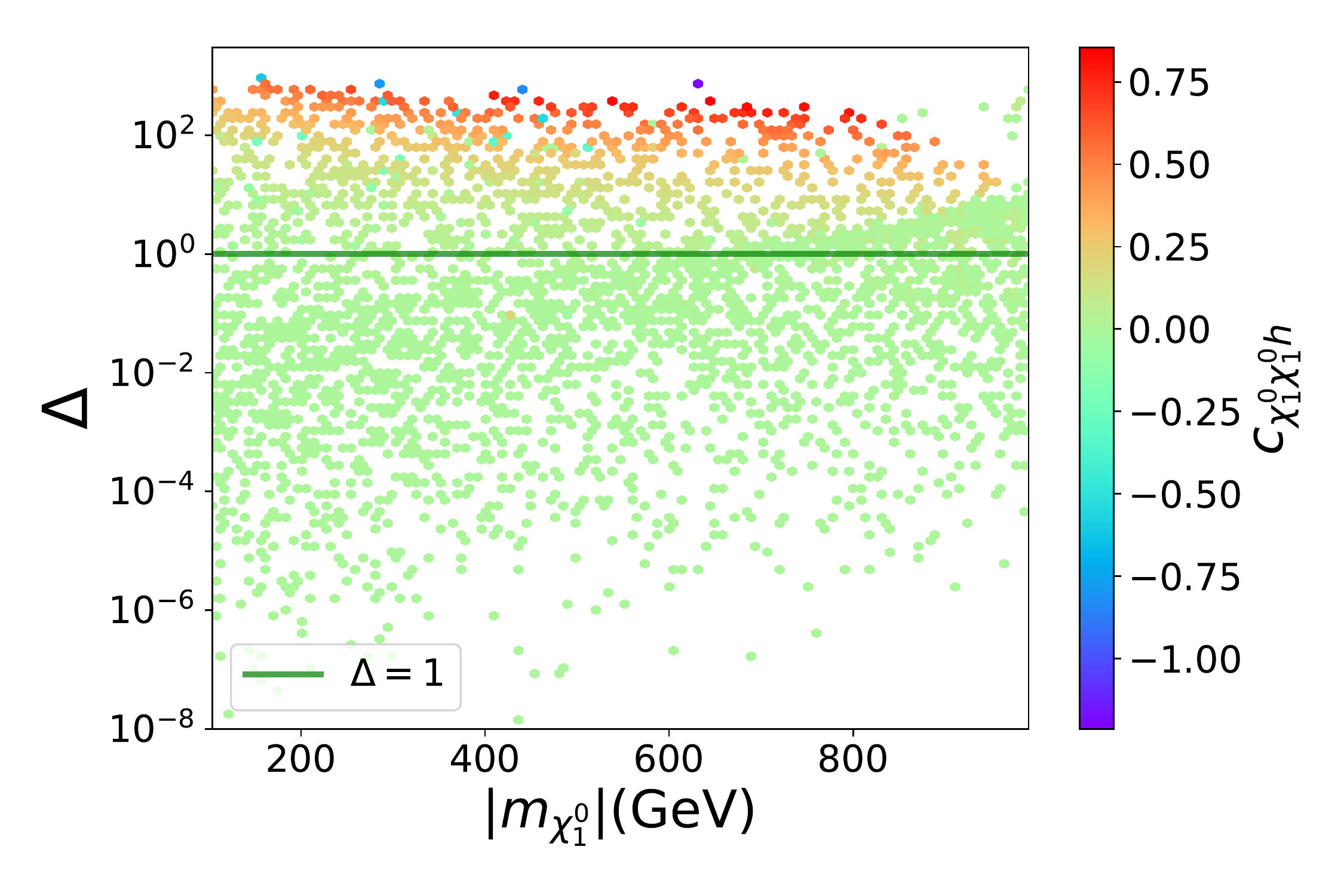}
	\caption{DD detection limits on the $\Delta$ parameter vs. $|m_{\chi_1^0}|$. The color gradient represents the fermion DM coupling to the Higgs and the green solid line is $\Delta=1.0$}
	\label{fig:DTF_DD}
\end{figure}

In Fig. \ref{fig:DTF_DD} the DD results are presented in the $\Delta$-$|m_{\chi_1^0}|$ plane with the $c_{\chi_1^0 \chi_1^0 h}$ in the colorbar, where all points shown satisfy the relic density constraint. The green solid line corresponds to $\Delta=1$, thus, all points above it are excluded by XENON1T~\cite{Aprile:2019dbj}. We thus find that $c_{\chi_1^0 \chi_1^0 h}$ must be less than 0.08 in order to still be viable. Moreover, we find that about 50 percent of all the points are excluded by DD. We do not present a similar plot to Fig. \ref{fig:DTF_DD} including scalar parameters becuase no additional constraints were found in that sector.

%%%%%%%%%%%% FIGURE%%%%%%%%%%%%%%%%%
%%%%%%%%%%%% FIGURE%%%%%%%%%%%%%%%%%

%%%%%%%%%% INDIRECT DETECTION %%%%%%%%%%%%%%
\subsection{Indirect detection}
\label{ID-DTF}
\begin{figure}[h]
	\centering
	\includegraphics[scale=0.38]{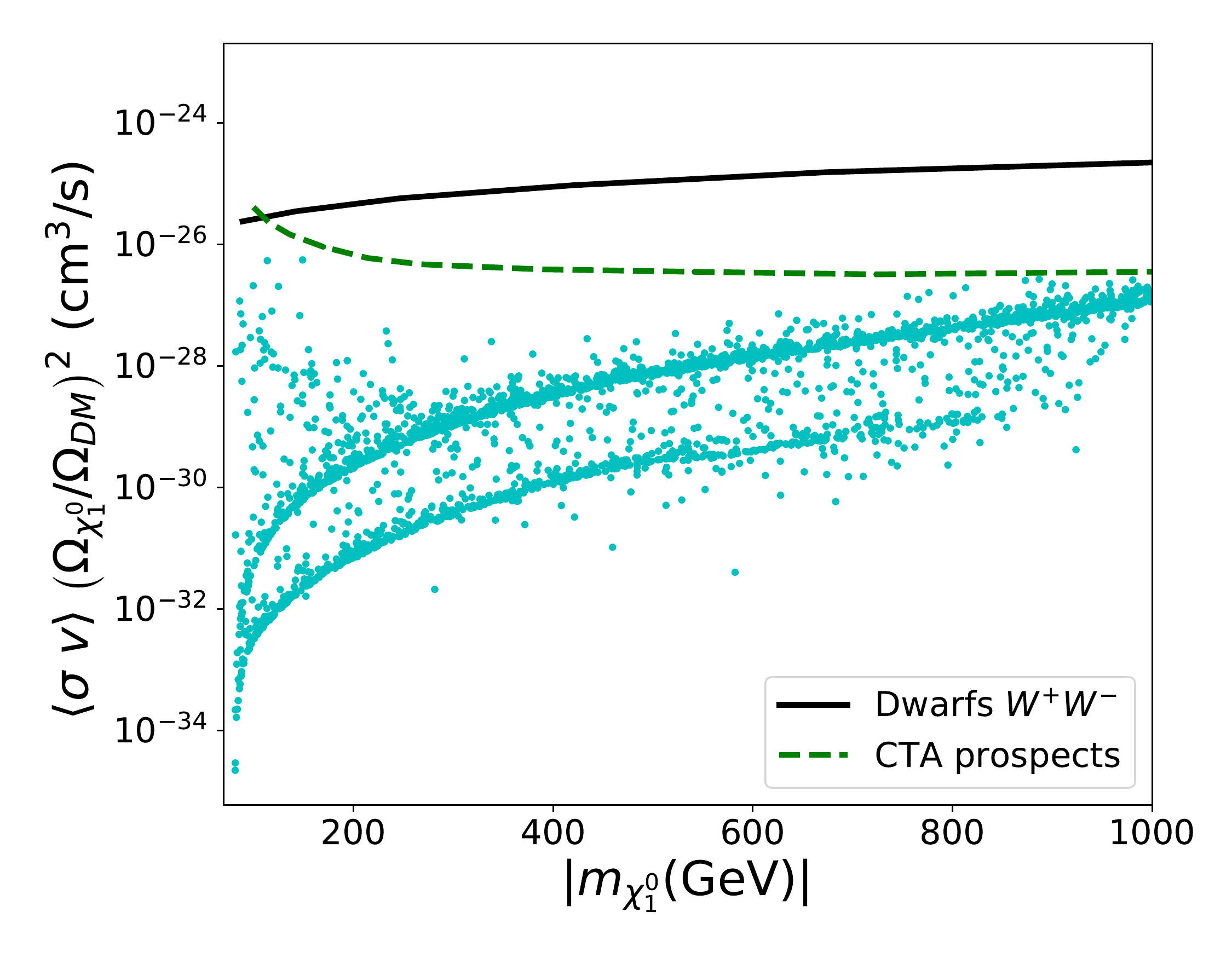}
	\includegraphics[scale=0.38]{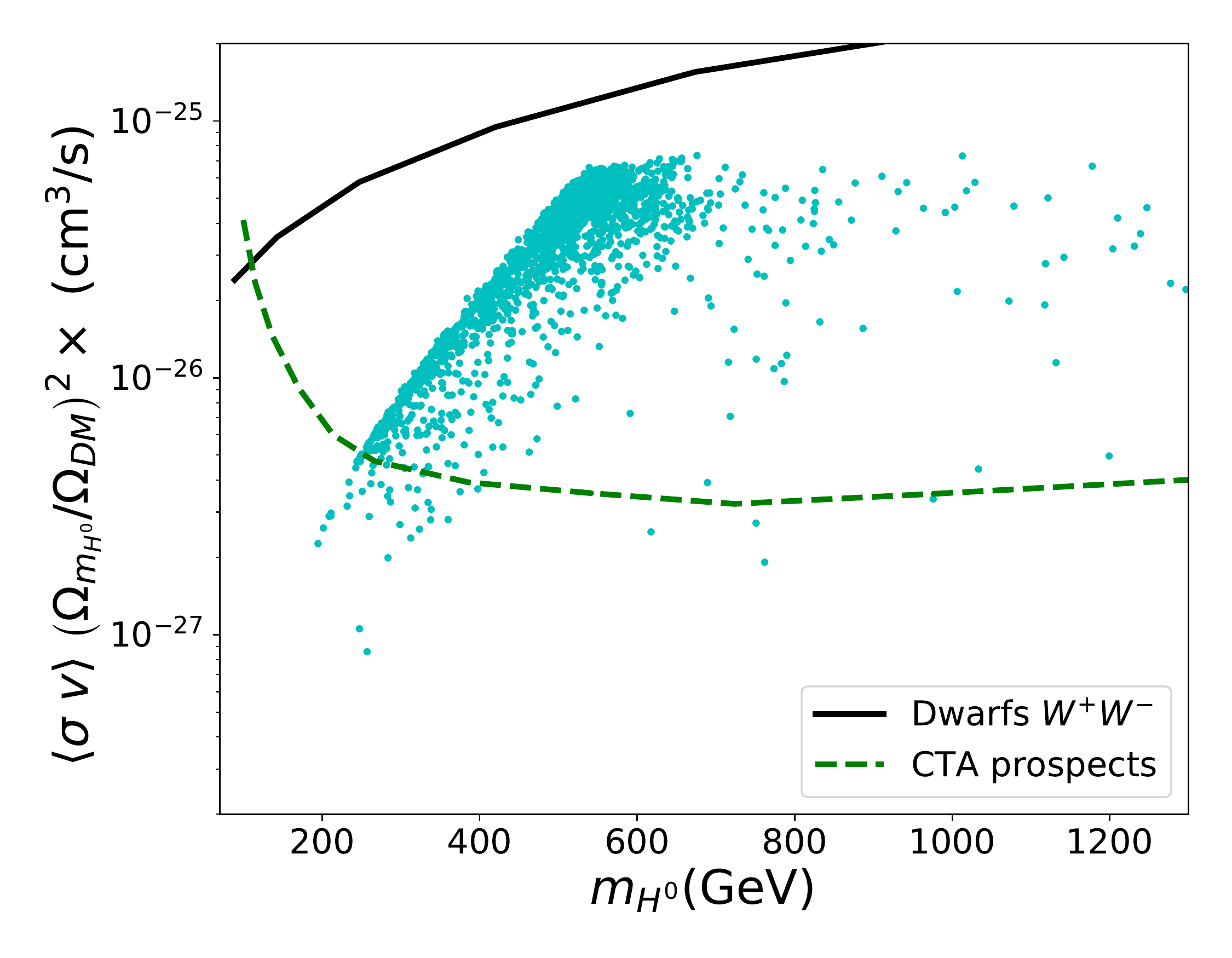}
	\caption{Left panel: rescaled fermion $\langle \sigma v \rangle$  vs. $|m_{\chi_1^0}|$. Right panel: rescaled scalar $\langle \sigma v \rangle$   vs. $m_{H^0}$.}
	\label{fig:DTFDM_S_ID}
\end{figure}

In regions where a high DM density is expected, such as dwarf spheroidal galaxies (dSphs) and the center the Milky Way, DM particles may find each other and annihilate into SM particles. The product of that annihilation may be visible as an excess, such as one in the gamma ray spectrum. The Fermi satellite searches for such gamma rays in dSphs and so far has found no deviations from the expected spectrum, thus, it imposes constraints on the thermally averaged DM annihilation cross section \cite{Ackermann:2015zua}.
In the case of multicomponent DM, the restrictions imposed by this observable are weakened, the reason is that, like DD, the event rate is dependent on the DM candidate local density. However, unlike DD, the event rate must be rescaled as $\left(\Omega_i / \Omega_{DM} \right)^2$, thus, a further suppression and loosened restrictions are expected. In fact, we found that current restrictions from the Fermi satellite (solid black curve) are well above the rescaled $\langle \sigma v \rangle$ for both the fermionic sector and scalar. Nevertheless, we present the prospects from the CTA experiment (green dashed curve) as given in~\cite{Funk:2013gxa,Wood:2013taa}. In the scalar sector most models will be explored by this experiment, whereas the fermionic content is out of reach. All of these results are presented in Fig.~\ref{fig:DTFDM_S_ID}.

%%%%%%%%%%%% FIGURE%%%%%%%%%%%%%%%%%
%%%%%%%%%%%% FIGURE%%%%%%%%%%%%%%%%%

%%%%%%%%%% COLLIDER %%%%%%%%%%%%%%
\subsection{Collider phenomenology}
\label{collider_DTF}

Due to the electroweak scale masses of the two DM candidates, it is, in principle, possible to produce them at the energies within reach of the LHC. The ATLAS and CMS collaborations look for signatures of such processes, with current analyses being consistent with the background only hypothesis. Thus, it is possible to place further restrictions on the model. Due to the imposed symmetries that guarantee the DM stability, we expect the fermion sector to be produced in separate processes than the scalar sector. This is actually a way multicomponent dark sectors can be explored. 
The fermion content of this model resembles that of the Wino-Higgsino model in the MSSM, thus, we may use the results from SUSY searches at the LHC. The limits are dependent on the processes and the mass splitting between the lightest charged fermion and the fermionic DM. For the region where  $m_{\chi_1^{\pm}} - m_{\chi_1^0} \geq 80.0$ GeV  we may use the results for searches where $p \ p \rightarrow \chi_1^{+} \ \chi_1^{-}$, $\chi_1^{\pm} \rightarrow  \chi_1^0 \ W^{\pm} \rightarrow \chi_1^{\pm} \nu \chi_1^0$. The ATLAs collaboration has presented exclusion limits for $\sqrt{s}=13$ TeV and 139~$\mathrm{fb^{-1}}$ in \cite{ATLAS:2019cfv}. Those limits are for the case when $\chi_1^{\pm}$ is Wino. In the case of the DTF model, the production cross section of viable models where $m_{\chi_1^{\pm}} - m_{\chi_1^0} \geq 80.0$ GeV resembles that of the Higgsino, thus, the exclusion limits are less constraining. However, after recasting the ATLAS exclusion limits, we find no additional constraints in the model. This happens because the Higgs diphoton decay rate places stronger constraints than the SUSY searches results from the ATLAS experiment. 
On the other hand, for models with mass splitting between 2 GeV $\leq m_{\chi_1^{\pm}} - m_{\chi_1^0} \leq 30$ GeV, the production cross section is also Higgsino, and though there are searches for that mass splitting such as the so-called compressed spectra, it is not possible to directly recast them, since they either correspond to the Wino case or to the Higgsino case with a very specific mass spectra. 
For the DTFDM model, the most common mass splitting lies between $m_{\chi_1^{\pm}} - m_{\chi_1^0} <$ 0.5 GeV, in that case, restrictions on long lived particles may apply which are the same as the ones described in the SDFDM model.

%%%%%%%%%% BP %%%%%%%%%%%%%%
\subsection{Benchmark points}
\label{BP_DTF}
In this section we include two benchmark points of the model that satisfy the constrains metioned in the previous sections:

\begin{center}
	\begin{tabular}{|l |l| l| r|}
		
		\hline
		
		BP &	Scalar Parameters    & Fermion Parameters & Observables \\
		\hline
		
		BP1&	$\mu_2=235.4 $ GeV      & $M_{\psi}=988$ GeV   & $\Omega_{\chi_1^0} h^2=0.098$      \\
		& $\lambda_3=$1.2$\times 10^{-3}$        & $M_{\Sigma}=-1040$ GeV & $\Omega_{H^0} h^2=0.024$      \\
		&	$\lambda_4=-$-4.4$\times 10^{-3}$       & $y_1=$6.5$ \times 10^{-2}$     & $\Delta=$1.6$\times 10^{-2
		}$     \\
		& $\lambda_5=$-1.5$\times 10^{-3}$ & $y_2=$9.0 $\times 10^{-4 	}$ & \\
		\hline
		\hline
		BP2& $\mu_2=382 $ GeV   & $M_{\psi}=844$ GeV     & $\Omega_{\chi_1^0} h^2=0.067$      \\
		& $\lambda_3=$-9.6$ \times 10^{-3}$        & $M_{\Sigma}=$-986 GeV & $\Omega_{H^0} h^2=0.051$      \\
		&	$\lambda_4=$-3.6$\times 10^{-2}$       & $y_1=$2.3 $\times 10^{-1}$     & $\Delta=$0.29     \\
		& $\lambda_5=$-5$\times 10^{-4}$ & $y_2=$7.4$ \times 10^{-2}$ & \\
		\hline	
	\end{tabular}
\end{center}

\section{singlet-triplet fermion dark matter model}
\label{sec:ST-model}
The singlet-triplet fermion DM model (STFDM model for short),
is an extension of the SM with  additional particle content:
i) A complex scalar doublet of $SU(2)_L$ $\eta$
which is odd under a discrete $Z_2$ symmetry. ii) Two hyperchargeless fermions;
a singlet $N$, and a  triplet $\Sigma$, of $SU(2)_L$ which are odd under a discrete
$Z^{\prime}_2$ symmetry. iii) A real scalar triplet $\Omega$ is also introduced to the model,
and this one as well as  the whole SM particle content are even under both discrete symmetries.
The STFDM model has been widely
studied in Ref.~\cite{Hirsch:2013ola, Merle:2016scw, Restrepo:2019ilz, Avila:2019hhv}.
The triplets in the standard $2 \times 2$ matrix notation of $SU(2)_L$ reads:

\begin{eqnarray}\label{eq:STFDM_triplets_decomposition}
\Sigma =
\begin{pmatrix} 
\frac{\Sigma^{0}}{\sqrt{2}} & \Sigma^{+} \\
\Sigma^{-} & -\frac{\Sigma^{0}}{\sqrt{2}}
\end{pmatrix}, \hspace{2cm}
\Omega =
\begin{pmatrix} 
\frac{\Omega^{0}}{\sqrt{2}} & \Omega^{+} \\
\Omega^{-} & -\frac{\Omega^{0}}{\sqrt{2}}
\end{pmatrix}.
\end{eqnarray}

The additional scalar doublet $\eta$ is decomposed as,
$\eta^{T}= \Big (\eta^{+}, \frac{1}{\sqrt{2}}  (  \eta^{R} + i \eta^{I} )    \Big )$.
The particle content of the model is
displayed in table~\ref{tab:particles_STFDM_1}.

%%%%%%%%%%%%%%%%%%%%%%%%%%%%%%%%%%%%%%%%%%%%%%%%%%%%%%%%%%%%%%%%%%%%%%%%%%%%
\begin{table}[h!]
\label{tab:particles_STFDM_1}
\begin{center}
\begin{tabular}{|c||c|c|c|c|c|c|}
\hline 
&  $SU(2)_{L}$ & $  U(1)_{Y}$   & $Z_2$ & $Z_{2}^{\prime}$ & $S$
\\ \hline \hline 
 $H$    &    $2$ & $1$ & $+$ & $+$ & $0$ \\
\hline
 $\eta$    &    $2$ & $1$ & $-$ & $+$ & $0$ \\
\hline
 $\Omega$  &    $3$ & $0$ & $+$ & $+$ & $0$ \\
\hline
 $N$         &  $1$ & $0$  & $+$ & $-$ & $1/2$ \\ 
\hline
 $\Sigma$    & $3$ & $0$   & $+$ & $-$ & $1/2$ \\ 
\hline
\end{tabular}
 \caption{ Quantum numbers of the particle
    content of STFDM  model under $SU(2)_L \otimes U(1)_Y \otimes Z_2 \otimes {Z_{2}^{\prime}}$.}

\end{center}
\end{table}
%%%%%%%%%%%%%%%%%%%%%%%%%%%%%%%%%%%%%%%%%%%%%%%%%%%%%%%%%%%%%%%%%%%%%%%%%%%%
The most general Lagrangian, invariant under  $SU(2)_L \otimes U(1)_Y \otimes Z_2 \otimes {Z_{2}^{\prime}}$
and involving the new fields takes the form~\cite{Hirsch:2013ola, Merle:2016scw, Restrepo:2019ilz, Avila:2019hhv}:
%%%%%%%%%%%%%%%%%%%%%%%%%%%%%%%%%%%%%%
\begin{align}
\label{eq:STFDM_Lagrangian}
\mathcal{L} = &\ 
   \mathcal{L}_{\rm{SM}} +  i {\rm Tr}\left[\overline{\Sigma} \slashed D\Sigma\right] 
   - \dfrac{1}{2}{\rm Tr}\left[
     \overline{\Sigma} M_{\Sigma} \Sigma^{c} + \overline{\Sigma^{c}} M_{\Sigma}^{*} \Sigma
                        \right] 
   - \left(      
     Y_{\Omega} \overline{\Sigma} \Omega N   + {\rm h.c.}
    \right) 
\nonumber\\ & + (D_\mu\eta)^\dagger(D^\mu\eta) + {\rm Tr}{(D_\mu\Omega)^\dagger(D^\mu\Omega)} - V (H,\eta,\Omega) 
\,,    
\end{align}  
%%%%%%%%%%%%%%%%%%%%%%%%%%%%%%%%%%%%%%

with
\begin{eqnarray}\label{eq:STFDM_Lagrangian}
V(H,\eta,\Omega)  &=& -\mu_{\phi}^{2} H^{\dagger} H - \mu_{\eta}^{2} \eta^{\dagger} \eta +
\dfrac{\lambda_1}{2} \Big ( H^{\dagger} H\Big )^{2} +   \dfrac{\lambda_2}{2} \Big ( \eta^{\dagger} \eta\Big )^{2} +
\lambda_3 \Big ( H^{\dagger} H\Big ) \Big ( \eta^{\dagger} \eta\Big )  \nn  \\
&+& \lambda_4 \Big ( H^{\dagger} \eta\Big ) \Big ( \eta^{\dagger} H\Big )  +
\dfrac{\lambda_5}{2} \Bigg [  \Big( H^{\dagger} \eta \Big )^{2} + \rm{h.c} \Bigg ] - \dfrac{m_{\Omega}^{2}}{2} \Omega^{\dagger} \Omega \nn \\
&+& \dfrac{\lambda_{1}^{\Omega}}{2} \Big ( H^{\dagger}  H \Big ) \Big ( \Omega^{\dagger}  \Omega \Big )
+ \dfrac{\lambda_{2}^{\Omega}}{4} \Big ( \Omega^{\dagger}  \Omega \Big )^{2}
+ \dfrac{\lambda^{\eta}}{2} \Big ( \eta^{\dagger}  \eta \Big ) \Big ( \Omega^{\dagger}  \Omega \Big ) \nn \\
&+& \mu_1 H^{\dagger} \Omega H + \mu_2 \eta^{\dagger} \Omega \eta~. 
\end{eqnarray}

After EWSB, the scalar fields develop VEV

\begin{eqnarray}\label{eq:STFDM_vev_scalar_sector}
  \langle H \rangle =
\begin{pmatrix} 
0  \\
\dfrac{v}{\sqrt{2}} 
\end{pmatrix}, \hspace{2cm}
\langle\Omega \rangle =
\begin{pmatrix} 
\dfrac{v_{\Omega}}{\sqrt{2}} & 0 \\
0 & -\dfrac{v_{\Omega}}{\sqrt{2}}
\end{pmatrix}.
\end{eqnarray}

Also, the Yukawa interaction mixes  $N$ and the neutral component of $\Sigma$ field,
with mass matrix:
\begin{eqnarray}{\label{eq:spectrum_fermion}}
M_{\chi}=\begin{pmatrix} 
M_{\rm{N}} & Y_{\Omega} v_{\Omega} \\
Y_{\Omega} v_{\Omega} & M_{\Sigma}
\end{pmatrix}~,
\end{eqnarray}

  and the physical states are obtained by the diagonalization of a $2 \times 2$ matrix,
  which is written in terms of the angle $\alpha$, such as:

\begin{eqnarray}{\label{eq:stfdm_fermion_mixing}}
\begin{pmatrix} 
\chi_{1}^{0}  \\
\chi_{2}^{0}
\end{pmatrix}
  =\begin{pmatrix} 
\cos\alpha & \sin\alpha \\
-\sin\alpha & \cos\alpha 
  \end{pmatrix}
\begin{pmatrix} 
\Sigma^{0}  \\
N
\end{pmatrix}~.
\end{eqnarray}  
  
Where the mixing angle $\alpha$  obeys:

\begin{eqnarray}{\label{eq:stfdm_mixing_angle_fermions}}
\tan (2\alpha) = \dfrac{2Y_{\Omega} v_{\Omega}}{M_{\Sigma} - M_{N}},
\end{eqnarray}

and the tree level fermion masses reads~\cite{Hirsch:2013ola, Merle:2016scw, Restrepo:2019ilz}:

\begin{eqnarray}{\label{eq:spectrum_fermions}}
  m_{\chi^{\pm}} &=& M_{\Sigma}~, \nn \\
  m_{\chi_1^{0}} &=& \dfrac{1}{2} \Bigg ( M_N + M_{\Sigma} - \sqrt{(M_{\Sigma} - M_N )^2  +4 Y_{\Omega}^2v_{\Omega}^2  }\Bigg)~, \nn \\
  m_{\chi_2^{0}} &=& \dfrac{1}{2} \Bigg ( M_N + M_{\Sigma} + \sqrt{(M_{\Sigma} - M_N )^2  +4 Y_{\Omega}^2v_{\Omega}^2  }\Bigg)~.
\end{eqnarray}

  Following explicitly the description of the STFDM model in Refs.~\cite{Merle:2016scw, Restrepo:2019ilz},
  we briefly describe the scalar spectrum of the model.
  \begin{enumerate}[label=\roman*]  
  \item) Firstly, the CP-even sector,
    in which the physical states $\Omega^{0}$ and $h$ get mixed,
     the $2 \times 2$  mixing matrix can be parametrized in terms of an angle $\beta$,
%%%%%%%%%%%%%%%%%%%%%%%%%%%%%%%%%%%%%%%%%%%%%%%%%%%%%%%%%%%%%%%%%%%%%%
\begin{eqnarray}{\label{eq:stfdm_scalar_mixing}}
\begin{pmatrix} 
h_1  \\
h_2
\end{pmatrix}
  =\begin{pmatrix} 
\cos\beta & \sin\beta \\
-\sin\beta & \cos\beta 
  \end{pmatrix}
\begin{pmatrix} 
h  \\
\Omega^{0}
\end{pmatrix}~.
\end{eqnarray}  
%%%%%%%%%%%%%%%%%%%%%%%%%%%%%%%%%%%%%%%%%%%%%%%%%%%%%%%%%%%%%%%%%%%%%%      
After EWSB,  there are two neutral states $h_1$ and $h_2$, the first one is identified as
  the observed Higgs field with a mass $m_{h_1}=125.09~\rm{GeV}$~\cite{Aad:2015zhl},
  and the second one corresponds to heavier electrically  neutral CP-even  scalar yet to be discovered.
  There is also a mixing between the states $\Omega^{+}$ and $H^{+}$, which after
  EWSB, transform
  into two electrically charged states, the first one becomes the longitudinal degree of freedom for
  the $W$ boson and the second one remains as a charged scalar $h^+$.
  From the scalar sector, $\eta^{R}$ is  chosen as lightest scalar,
  charged under $Z_2$ and  stands as the scalar DM
  candidate with a tree level mass:

  \begin{eqnarray}{\label{eq:stfdm_scalar_spectrum_1}}
    m_{\eta^{R}}^2 &=&  \mu_{\eta}^2 + \dfrac{1}{2}(\lambda_3 + \lambda_4 + \lambda_5)v^{2} +
    \dfrac{1}{2}\lambda^n v_{\Omega}^2 - \dfrac{1}{\sqrt{2}}v_{\Omega}\mu_2~. 
\end{eqnarray}

\item) Secondly, for CP-odd sector there are not
  mixing and the fields  $\eta^{I}$ and $\eta^{\pm}$ acquire masses.

  \begin{eqnarray}{\label{eq:stfdm_scalar_spectrum_2}}
    m_{\eta^{I}}^2 &=&  \mu_{\eta}^2 + \dfrac{1}{2}(\lambda_3 + \lambda_4 - \lambda_5)v^{2} +
    \dfrac{1}{2}\lambda^n v_{\Omega}^2 - \dfrac{1}{\sqrt{2}}v_{\Omega}\mu_2~,  \nn \\
    m_{\eta^{\pm}}^2 &=&  \mu_{\eta}^2 + \dfrac{1}{2}\lambda_3 v^{2} +
 \dfrac{1}{2}\lambda^n v_{\Omega}^2 +\dfrac{1}{\sqrt{2}}v_{\Omega}\mu_2~. 
\end{eqnarray}  

 \end{enumerate}

 Note that the origin of neutrino mixing and  masses can not be explained within the context of the STFDM model
due to the imposed discrete $Z_2 \times Z_{2}^{\prime}$  symmetry which guarantees the co-existence of the two DM species.

%start of DM conversion section
\subsection{DM conversion in the STFDM model}
\label{DM_conversion_stfdm}
{The coexistence of the two DM  species--
  the fermionic and scalar-- in the early  Universe
  allows them to transform into each other.
  Considering the limit in which
  the real scalar triplet $\Omega$ is decoupled\footnote{Even though,
  the scalar field $\Omega$ is allowed to develop a non-zero VEV.},
  the Higgs portal is the one  connecting
  the two DM sectors. Such a conversion is controlled mainly
  by two parameter,   $\lambda_L = (\lambda_3+ \lambda_4+\lambda_5)/2$,
  which connects the scalar DM to the Higgs 
  and $Y_{\Omega}$, which is the connection of the fermionic DM to the
  Higgs. Fig.~\ref{fig:stf_dm_conversion}
  shows the relic density $\Omega h^{2}$ as a function of
  mass of scalar DM specie for the IDM (dashed blue line)
  as well as for the scalar DM specie of STFDM (solid green line).
  The plot is obtained fixing the next parameters:
  $Y_\Omega=1.3$, $\lambda_2=0.1$, $\lambda_3=0.1$, $\lambda_4=10^{-6}$,
  $\lambda_5=-0.01$, $\lambda^{n}=0$, $\lambda_{1}^{\Omega}=0$, $\lambda_{2}^{\Omega}=0$,
  $\mu_2=0$,  $\mu_1=1000~\rm{GeV}$ and  $M_{\Sigma}=100~M_N$.
  In the  scenario under consideration, a scan in $M_N$ is performed in such
  a way that, $|m_{\chi_{1}^{0}} - m_{\eta^{R}}| \leq  0.05~\rm{GeV}$
  for each of  points displayed in the plot.
  The conversion process between the two DM candidates is
  described by the Feynman diagram displayed in Fig.~\ref{fig:Higgs_portal}.
  For the selected scan of the parameter space of the model, the Fig.~\ref{fig:stf_dm_conversion} 
  shows how the fermionic DM converts into scalar DM in the STFDM, increasing
  significantly its relic abundance.
    In the limit used for the example,
      the scalar field $\eta^R$ in the STFDM model is exactly $H^{0}$, the lightest
  neutral component of the IDM. It is worth
  to mention that the conversion process is most efficient
  in the way as the two DM sectors are almost mass degenerated.
  The $\Omega_{\chi} h^{2}$ is not
  shown in the figure since it is too large, it corresponds to a scenario
  in which $\chi_1^{0}$ is mostly singlet and therefore overabundant in
  the low mass regime under exploration. 
  The  results obtained for DM conversion are just an example that such a phenomena do happens in this model,
    however, is not phenomenological viable because the total relic density (  the contribution
  of both DM species )  is too large, and therefore excluded by current Planck satellite measurements.

\begin{figure}[t!]
\includegraphics[scale=0.38]{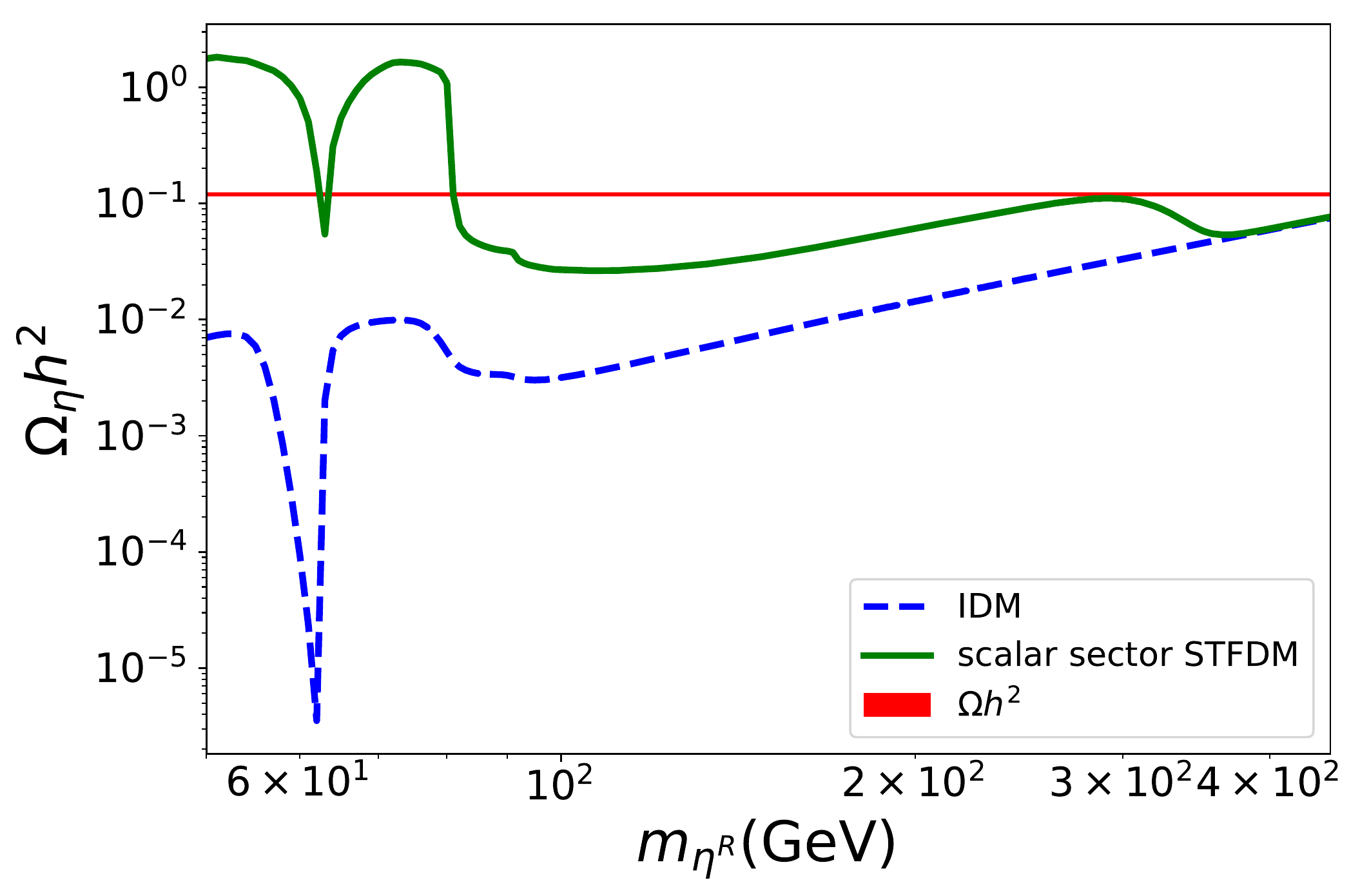}
\caption{
      DM conversion scenario in the STFDM model. Relic density as a function
   of the scalar DM mass. The dashed  blue  and solid green lines represents
   the $\Omega h^2$  for the IDM and the  STFDM model respectively.  The red
   band stands for the 3$\sigma$ observed relic density by Planck satellite.}
{\label{fig:stf_dm_conversion}}
\end{figure}

}
%% end of DM conversion section

\subsection{Numerical results}
\label{sec:pheno_stfdm}
As in the previous two models, the STFDM possesses two DM species,
the fermionic one $\chi_{1}^{0}$, which arise as the lightest component
of the   $N-\Sigma$  fermion mixing
and the lightest neutral scalar component of the $\eta$ doublet,
which is chosen to be the CP-even $\eta^{R}$\footnote{
  It is worth to  mention that the CP-odd  $\eta_{I}^{0}$ can also play
  the role of scalar DM, but the phenomenology in
  such a case does not differ too much from the one
obtained  by considering  $\eta_{R}^{0}$ instead.}.
% Little description of the model implementation
The model has been implemented in
% {\tt SARAH (v4.13.0)}~\cite{Staub:2013tta}
{\tt SARAH}~\cite{Staub:2013tta}
and then exported to
%{\tt micrOMEGAs (v5.0.4)}~\cite{Belanger:2018mqt},
{\tt micrOMEGAs}~\cite{Belanger:2018mqt},
where
dark matter observables, such as the  relic abundance, direct detection and indirect detection
were evaluated.
For the collider phenomenology and production cross section
computation, the model is exported in the  Universal FeynRules
Output (UFO)  format 
to the parton-level Monte Carlo (MC) generator
{\tt MadGraph (v5.2.5.5)}~\cite{Alwall:2014hca}. 
We carry out a scan in the parameter space of the model
described in Table~\ref{table:STFDM_parameter_space}.
\begin{table}
\centering
\begin{tabular}{|c|c|} 
\hline
Parameter & Range\\
\hline
$m_{N}$ & $1-5000$~(\rm{GeV})   \\
$m_{\Sigma}$ & $100-5000$~(\rm{GeV})  \\
%$\mu_{\eta}$ & $10-3000$~(\rm{GeV})   \\
$\mu_{i}$ & $10-5000$~(\rm{GeV})   \\
$Y_{\Omega}$ & $10^{-3}-3$   \\
%$\lambda_{2}$ &  $10^{-1}-10^{1}$   \\
%$|\lambda_{3,4,5 }|$ & $10^{-3}- 3$  \\
$|\lambda^{\eta}|$ & $10^{-3} - 3$   \\
$|\lambda_i^{\Omega}|$ & $10^{-3}- 3$  \\
\hline
\end{tabular}
\caption{Scan range of the parameters of
  the STFDM model. The $\mu_{\eta}$ and $\lambda_{i}$, for $i \in \lbrace 2,\ldots,  5 \rbrace$
  in the scalar sector are scanned as is shown in Table~\ref{table:IDM_scan}.}
\label{table:STFDM_parameter_space}
\end{table}

The VEV developed by the scalar triplet, is fixed to  $v_{\Omega}=5~{\rm{GeV}}$,
which is its possible maximum value allowed in order to fulfill the 
$\rho$ parameter
constraint~\cite{Tanabashi:2018oca, Merle:2016scw}.
All the simulated data  satisfy constraints of
perturbativity, the scalar potential is bounded from below,
LEP collider limits, Higgs diphoton decay rate, and EWPO
described in Sec.~\ref{sec:pheno_constraints}.
The contributions to the oblique S and T parameters
due to the additional field content of the model
is given in appendix~\ref{ewpo_stfdm}.
In the following subsection we describe the phenomenology of
the model.

\subsection{Relic Abundance}

\begin{figure}
\includegraphics[scale=0.38]{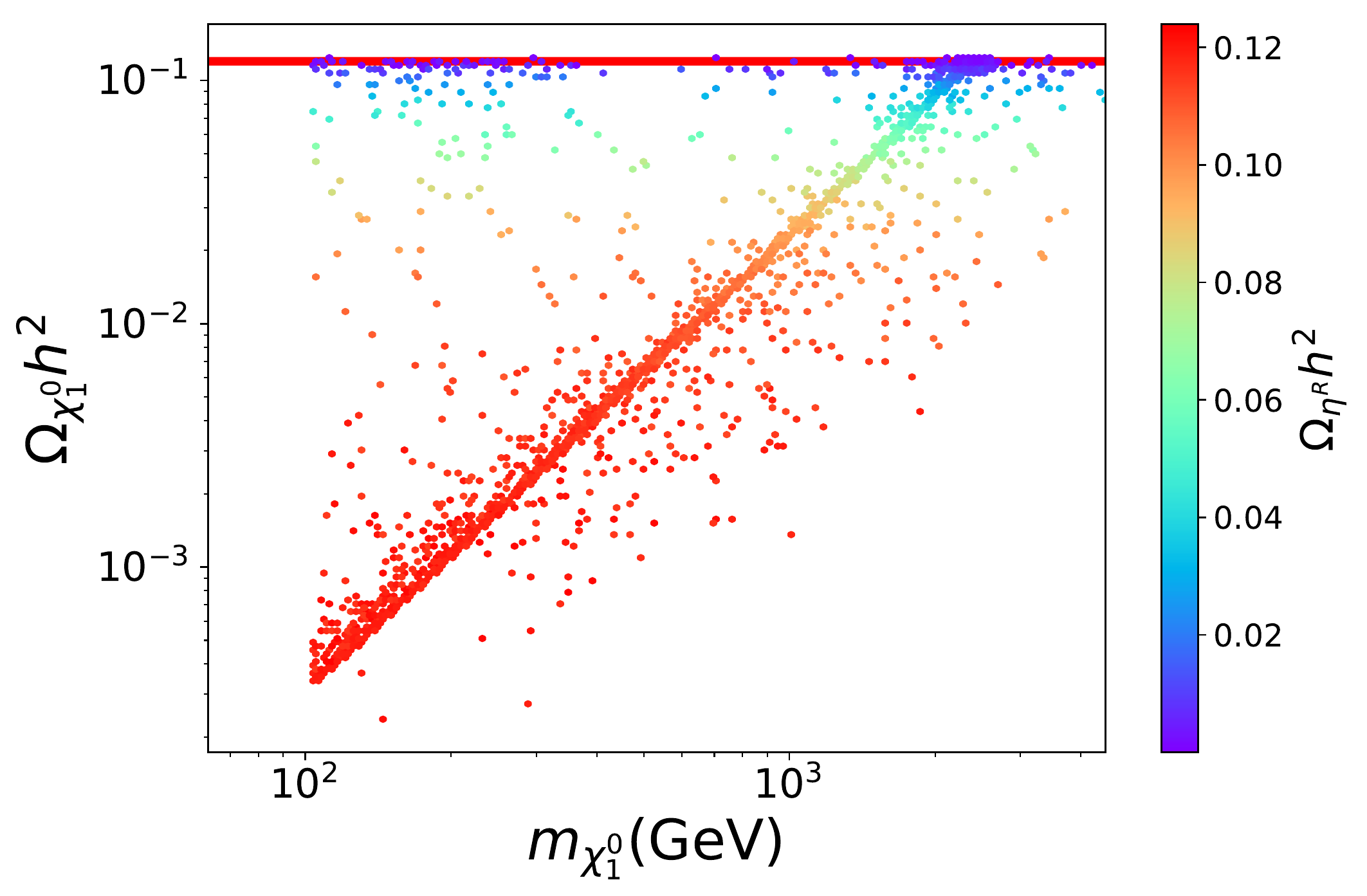}
\includegraphics[scale=0.38]{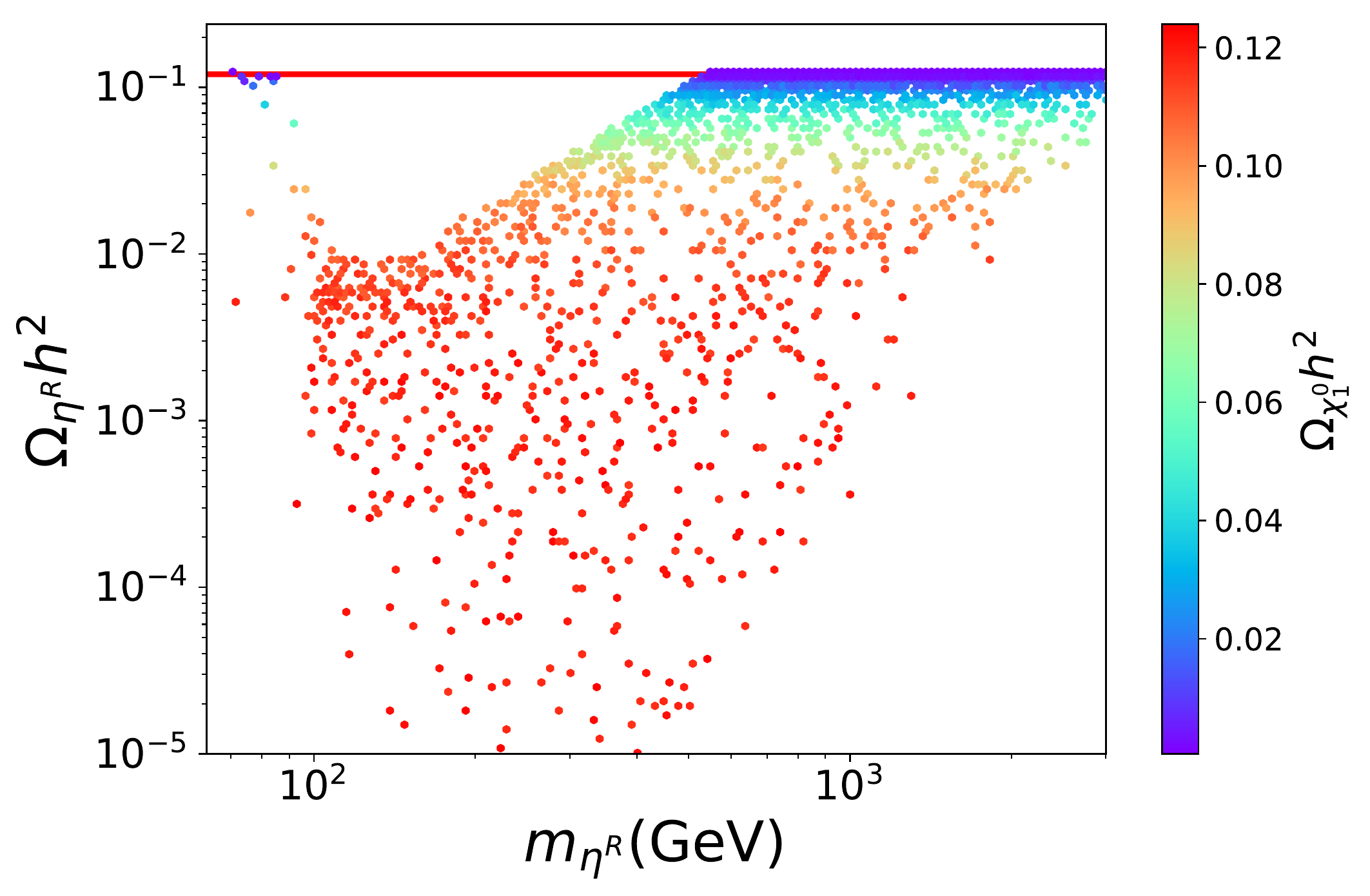}
\caption{Relic density $\Omega_i h^{2}$  as a function of
  the mass DM for each DM specie $i$, with $i \in \lbrace \chi_1^{0}, \eta^{R} \rbrace$.
  On the left (right), the
  plot shows the scenario for the fermionic (scalar) DM
  component. The most dense region on the left
  corresponds to the scenario in which the fermionic
DM candidate is mostly triplet.}
{\label{fig:relic_density_stf}}
\end{figure}

  In  Fig.~\ref{fig:relic_density_stf} the relic density
for the two DM species is displayed as a function
of their respective masses. In both plots all the points 
correspond to the full data set after imposing all the constraints
mentioned in Sec.~\ref{sec:pheno_stfdm}.
The most dense region on the left panel of the figure corresponds to
the case in which $\chi_1^{0}$ is mostly triplet,
this species alone can account for the $100\%$ of
the observed relic density when  $m_{\chi_{1}^{0}} \sim 2.5~\rm{TeV}$.
In the mass windows $ 100~{\rm{GeV}}< m_{\chi_{1}^{0}} < 2.5~{\rm{TeV}}$,
$\chi_1^{0}$ can completely explain the observed relic
density, thanks to the mixing of $N-\Sigma$. 
The color gradient shows the relic density
    associated to the scalar DM specie.   
On the right side of Fig.~\ref{fig:relic_density_stf},
the scalar DM can not account for the total relic abundance in the mass windows   
$ 100~{\rm{GeV}}< m_{\eta^{R}} < 550~{\rm{GeV}}$, this
due to the gauge interactions. On the other hand,  for  $m_{\eta^{R}} > 550~\rm{GeV}$
the scalar DM  alone can account for the total relic density.
 The color gradient shows
  the relic density  associated to the fermionic  DM specie.  
The red band in both plots correspond to the points with observed relic
density  at 3$\sigma$ CL.
With the interplay of the two DM sectors,
the total relic density is explained in
  the region
  $ 100~{\rm{GeV}}< m_{\rm{DM}} < 1.0~{\rm{TeV}}$.

%%%%%%%%%%%%%%%%%%%%%%%%%%%%%%%%%%%%%%%%%%%%%%%%%%%%%%%%%%%%%%      
\begin{figure}[t!]
\includegraphics[scale=0.38]{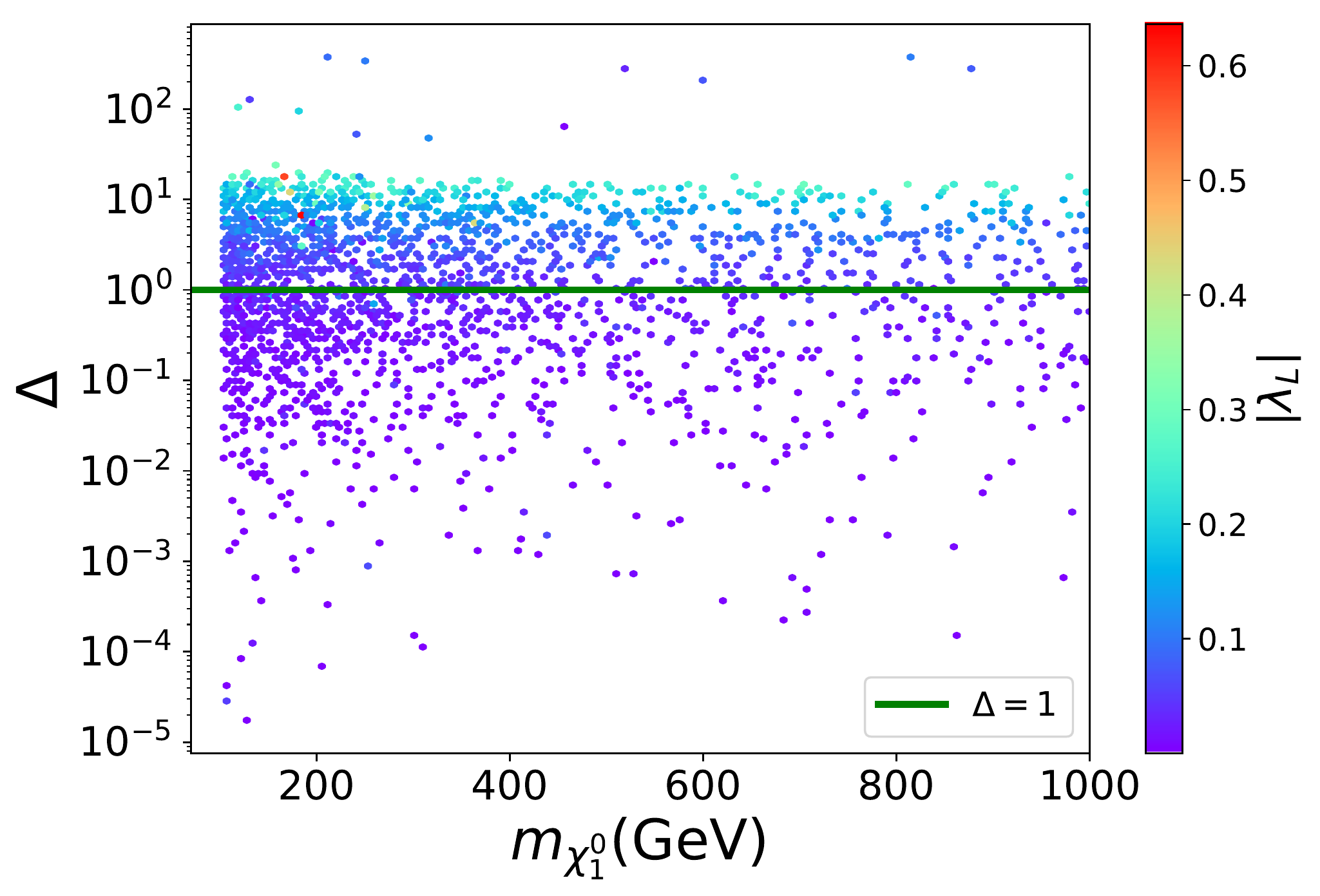}
\includegraphics[scale=0.38]{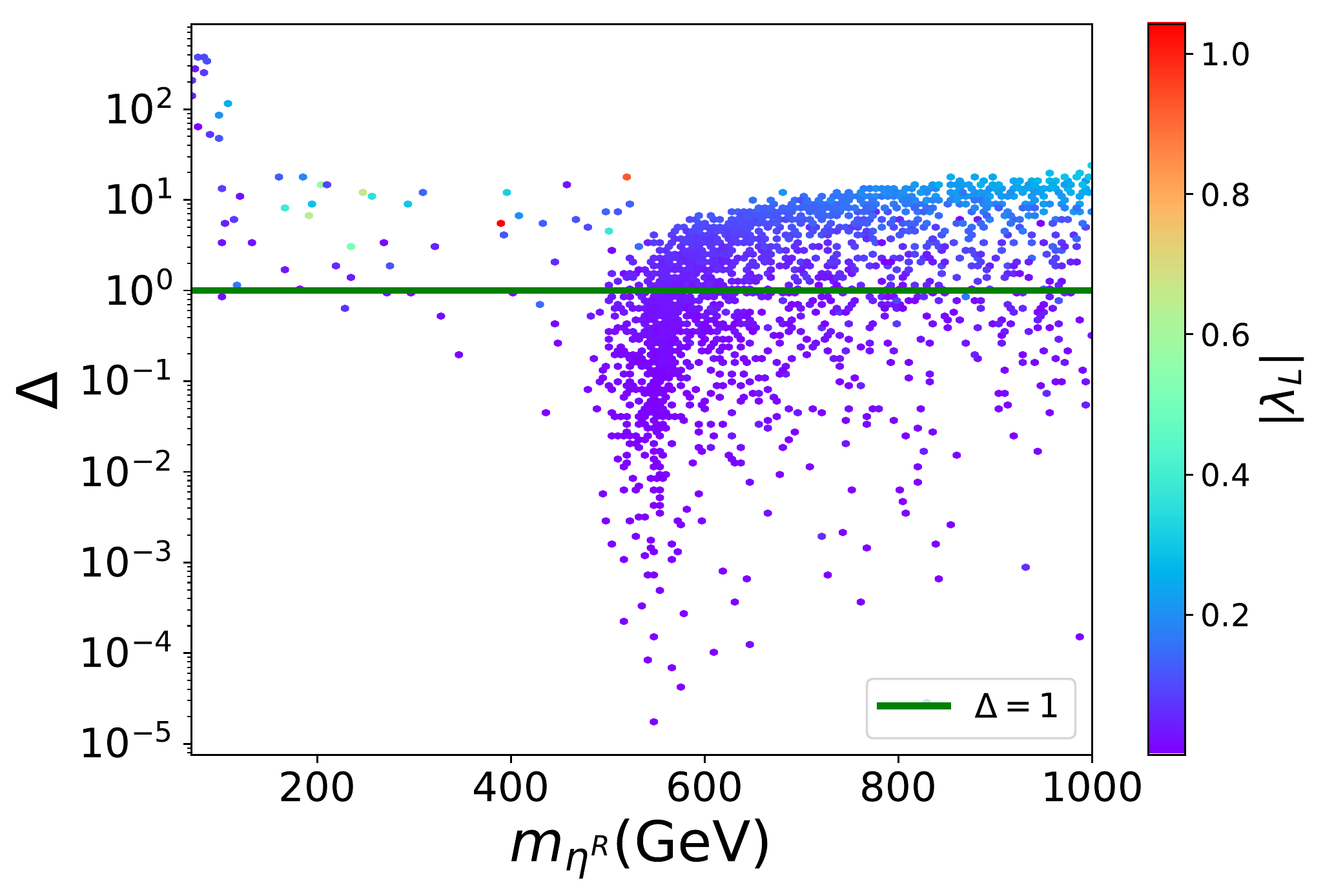}

 %% \caption{ Rescaled SI  cross section
 %%   for each specie of dark matter $\sigma_{i}^{\rm{SI}} \times \Omega_ih^{2} / \Omega_{DM}h^{2}$
 %%   as a function of its mass $m_i$, $i \in \lbrace \chi_1^{0}, \eta^{R} \rbrace$.
 %%   The green line represents the current  $2\sigma$ CL upper limit
 %%   on the cross section by XENON$1\rm{T}$, the black dashed line
 %%   stands for the future upper reach on SI cross section by DARWIN and
 %%   the neutrino coherent scattering background is given by the magenta region.
 %% }

\caption{ Value for $\Delta$  as a function of the dark
  matter mass for each specie,
  $m_i$, $ i \in \lbrace \chi_1^{0}, \eta^{R} \rbrace $. The green line ($\Delta = 1 $) represents
  the current upper limit on the $\Delta$ due to XENON$1\rm{T}$ restrictions. Points above
  the green light are ruled out.}
{\label{fig:sigma_si_stf}}
\end{figure}
%%%%%%%%%%%%%%%%%%%%%%%%%%%%%%%%%%%%%%%%%%%%%%%%%%%%%%%%%%%%%%  

\subsection{Direct detection}

  For the fermionic DM, the
  tree level SI cross section $ \sigma_{\chi_{1}^{0}}^{\rm{SI}}$ is given by~\cite {Merle:2016scw, Restrepo:2019ilz}:

  \begin{eqnarray}{\label{eq:stf_fermion_si_xs}}
    \sigma_{\chi_{1}^{0}}^{\rm{SI}}&=& \dfrac{\mu_{\rm{red}}^{2}}{ \pi}  \Bigg [ \dfrac{M_N f_N}{v} \dfrac{Y_{\Omega} \sin(2\alpha)\sin(2\beta)}{2}
      \Bigg ( \dfrac{1}{m_{h_2}^{2}}  - \dfrac{1}{m_{h_1}^{2}} \Bigg ) \Bigg ]^{2} ,
\end{eqnarray}

  where $f_N \approx 0.3 $  is the form factor for the scalar interaction~\cite{Alarcon:2011zs,Alarcon:2012nr}.
  $M_N \approx 0.938~\rm{GeV}$ the nucleon mass,
  $\mu_{\rm{red}}$ the reduced mass define as
  $\mu_{\rm{red}}=M_N m_{\chi_1^{0}}/ (M_N + m_{\chi_1^{0}} )$.
For DD limits, since for the two DM species the masses are larger
$\mathcal {O}(100~{\rm{GeV}} - 1~{\rm{TeV}})$, the recoil energy
spectra for all signals will have the same shape\footnote{As $m_{\rm{DM}} >> m_{Xe} $,
  the detector can not distinguish one DM candidate from the other.}. It allows to apply the constraint

\begin{eqnarray}
  \Delta = \dfrac{\sigma^{SI}_{\eta^R}}{\sigma^{SI}_{X_e}(m_{\eta^R})}
  \Bigg( \dfrac{\Omega_{\eta^R}}{\Omega} \Bigg ) + 
    \dfrac{\sigma^{SI}_{\chi_1^0}}{\sigma^{SI}_{X_e}(m_{\chi_1^0})}
    \Bigg( \dfrac{\Omega_{\chi_1^0}}{\Omega} \Bigg)  <1~.
\end{eqnarray}

In  Fig.~\ref{fig:sigma_si_stf}
is shown $\Delta$ as a
function of the mass of each DM specie and in color gradient is shown $|\lambda_L|$.
For the two panels, all the points with $|\lambda_L| \lesssim {\mathcal{O}}(0.2) $
are below the green line, and therefore allowed. On the right panel of
Fig.~\ref{fig:sigma_si_stf}, all points in the mass
windows $ 100~{\rm{GeV}}< m_{\rm{DM}} < 200~{\rm{GeV}}$ are ruled out.
For $m_{\eta^{R}} > 400~\rm{GeV}$, in the available parameter space, the scalar
mass spectra fulfills $m_{\eta^{+}} \sim m_{\eta^{R}} \sim m_{\eta^{I}} $,  due to
EWPO constraints.

\subsection{Collider phenomenology}

Following
the  criteria for the explanation of observed DM relic density in the mass windows
$100~{\rm{GeV}} <m_{\chi^{0}_{1}}< 1~{\rm{TeV}}$, the next general benchmark scenarios for
the collider analysis are defined:

\begin{enumerate}[label=\roman*]

\item {\bf A}: $m_{H^{+}} > m_{\chi^{+}_1} \approx m_{\chi^{0}_2} > (m_{\chi^{0}_1} + m_{W^{\pm}})~.  $

\item {\bf B}: $m_{H^{+}} >  m_{\chi^{+}_1}$  and    $5~{\rm{GeV}} <(m_{\chi^{+}} - m_{\chi^{0}_1}) <  50~{\rm{GeV}}~.$  

\item {\bf C}: $m_{H^{+}} >  m_{\chi^{+}_1}$  and    $m_{\pi^{\pm}} <(m_{\chi^{+}} - m_{\chi^{0}_1}) < 0.5~\rm{GeV}~.$  

\end{enumerate}

\begin{figure}[t!]
 \centering
\includegraphics[scale=0.38]{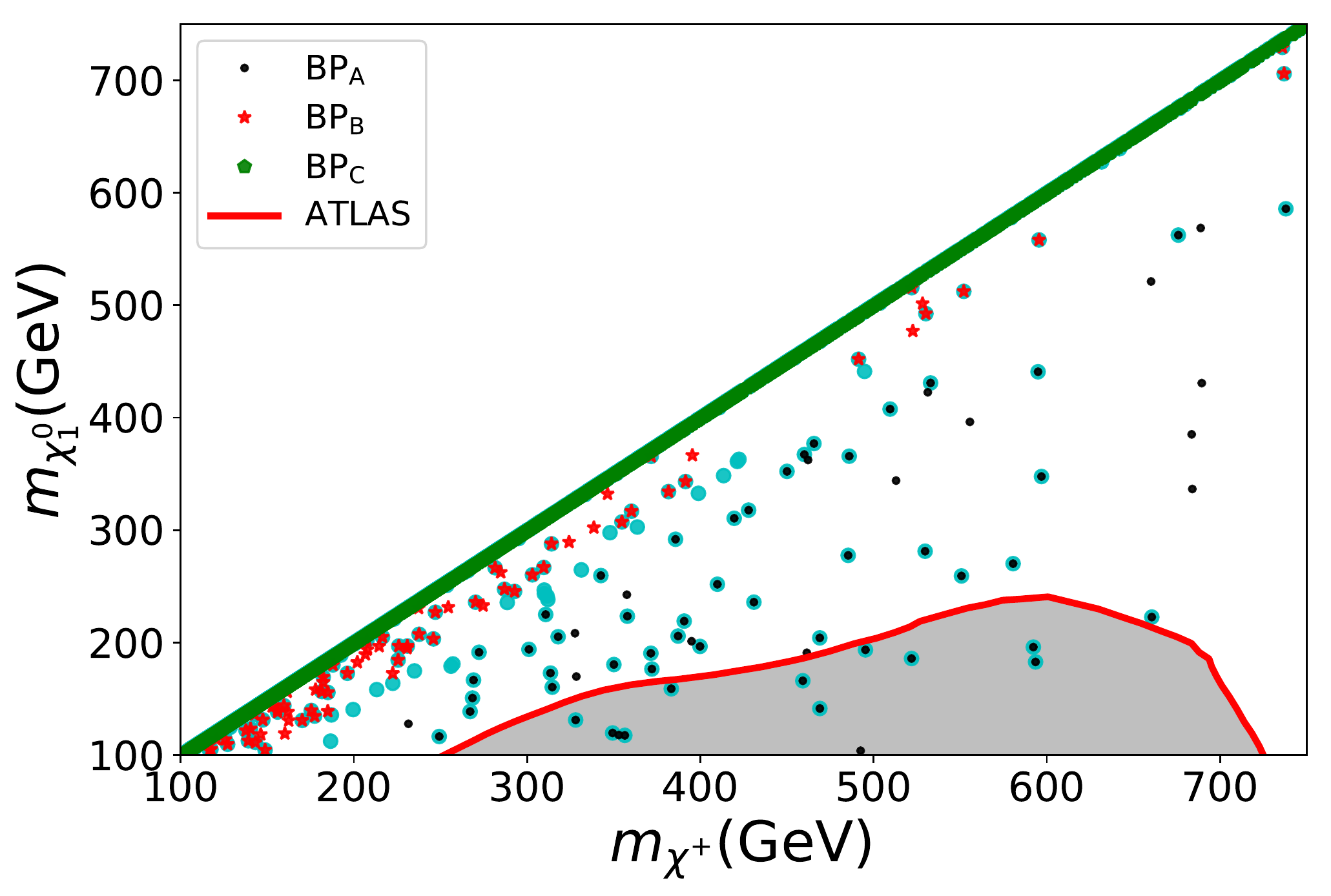}
\includegraphics[scale=0.38]{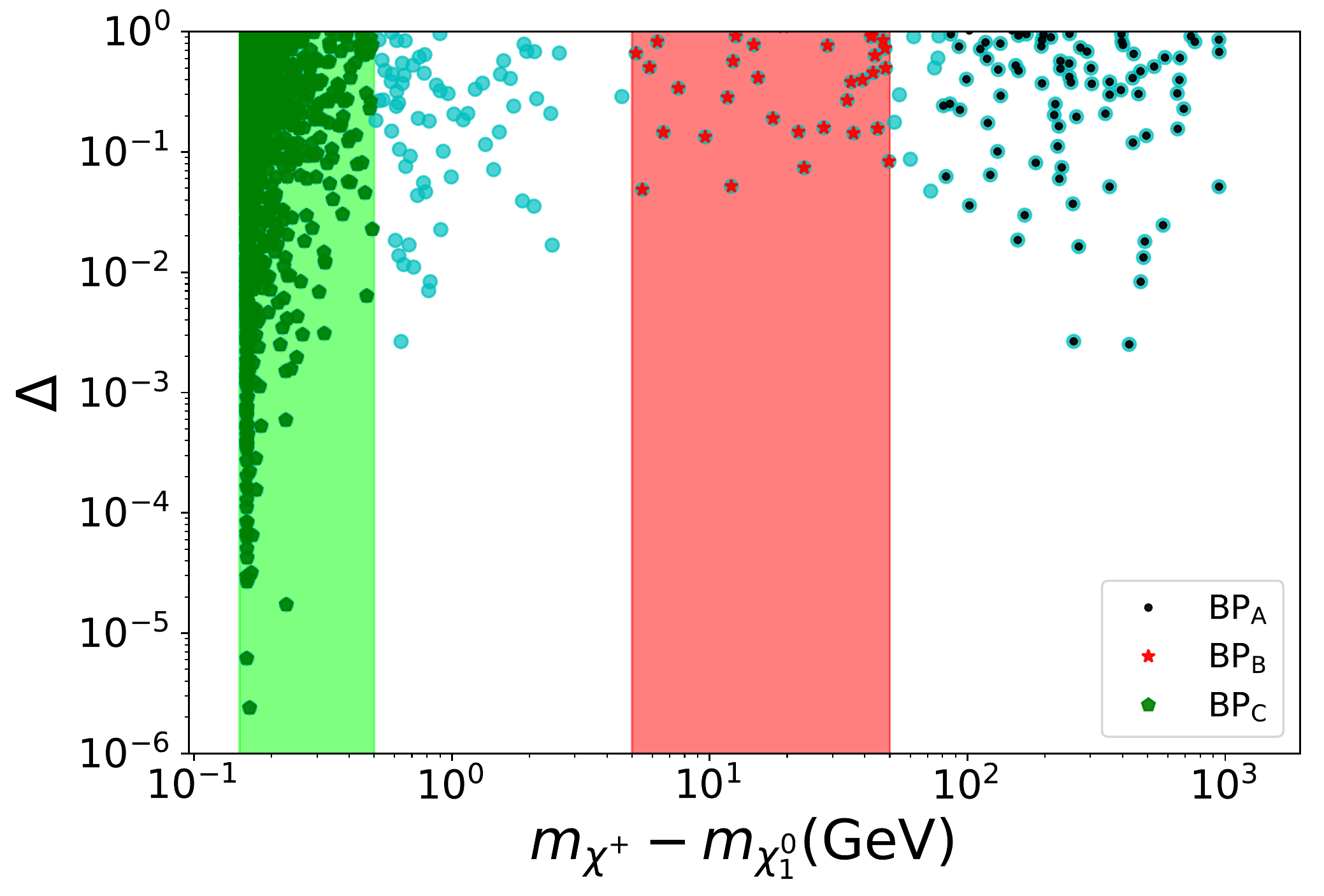}
\caption{ On the left panel, 
$m_{\chi_1^{0}}$ as a function of $m_{\chi^{+}}$, all the points
  displayed satisfy the observed relic density, direct detection and indirect detection
  current limits. The region in gray, which  correspond to points in which $m_{\chi^{+}}$ and $m_{\chi_2^{0}}$
  are mostly triplet and mass degenerate is currently ruled out by
  ATLAS searches~\cite{ATLAS:2019efx}. On the right panel, the $\Delta$ function
  vs $m_{\chi^{+}} - m_{\chi_1^{0}}$ plane is shown.
}
 {\label{fig:stf_collider_1}}
 \end{figure}

%%%% Discussion for scenario $A$
For scenario ${\bf A}$,  which correspond to the black points
  displayed in Fig.~\ref{fig:stf_collider_1}, the direct production of $\chi^{\pm} \chi_{2}^{0}$ at proton-proton collisions
is copiously since $\chi^{\pm}$ as well as $\chi_{2}^{0}$ are mostly triplet. The exclusion limit
in this case is settled following ATLAS results
for chargino-neutralino production from proton-proton collisions
at center of mass energy $\sqrt{s}=13~\rm{TeV}$ and with an
integrated luminosity of 139 $\rm{fb}^{-1}$~\cite{ATLAS:2019efx}.
In the STFDM, the process
$p \ p \to \chi^{\pm}\chi_2^{0}$,  ($\chi_1^{\pm} \to W^{\pm} \chi_{1}^{0}  \to  l^{\pm} \nu_L \chi_{1}^{0}$)
($\chi_2^{0} \to h_1 \chi_{1}^{0}  \to  b \overline{b} \chi_{1}^{0}$)
leading  to one  charged lepton (either electron or muon),
two b jets  and missing transverse energy ($E_{\rm{T}}^{\rm{miss}}$),
is exactly the one considered for the MSSM in the analyses of Ref.~\cite{ATLAS:2019efx}.
Since the production and decay are exactly the same of those of the MSSM
considered in one of the  ATLAS analyses, then, for fermionic DM with $m_{\chi^{0}_1} \approx 200~\rm{GeV}$,
the  fermion triplets (either $\chi^{+}, \chi^{0}_{2}$ ) are excluded up
to a mass of $m_{\chi^{+}} / m_{\chi^{0}_2} \approx 650~\rm{GeV}$.
All the excluded points by this analysis are shown  on the left panel
of Fig.~\ref{fig:stf_collider_1} and correspond to the ones
in the gray region and below the red line.

%%%% Discussion for scenario B
In scenario ${\bf B}$,  the mass interval
$5~\rm{GeV} <(m_{\chi^{+}} - m_{\chi^{0}_1}) < 50~\rm{GeV} $
correspond to a compressed mass spectra
and are the points in red displayed in
Fig.~\ref{fig:stf_collider_1}.
Such a compressed spectra scenarios are being study for simplified MSSM
in CMS through VBF production channels~\cite{Sirunyan:2019zfq}
and in ATLAS through s-channel production of charginos~\cite{ATLAS:2019lov},
in  both cases, the chargino decaying into neutralino and soft leptons.
The two analyses are complete, however the constraints does
not apply directly in the STFDM model, and a full
analysis is currently beyond scope of this work.
%%%% Discussion for scenario C
Scenario ${\bf C}$, is defined by the mass interval
$m_{\pi^{\pm}}  <(m_{\chi^{+}} - m_{\chi^{0}_1}) < 0.5~\rm{GeV} $,
with $m_{\pi^{\pm}}=139.6~\rm{MeV}$, the charged Pion mass. This
general benchmark correspond to the points
in green on Fig.~\ref{fig:stf_collider_1}.
In this case, for the mentioned mass
windows above,  the most predominant decay mode of charged fermion
is $\chi^{\pm} \to \pi^{\pm} \chi_1^{0} $, with $Br(\chi^{\pm} \to \pi^{\pm} \chi_1^{0} ) \geq 0.97 $,
however, the charged fermion $\chi^{\pm}$ have small width decay, allowing it to 
travel inside the detector before decay~\cite{Sirunyan:2018ldc}.
The width decay for the fermion  $\chi^{\pm}$ decaying to charged Pion reads:
%%%%%%%%%%%%%%%%%%%%%%%%%%%%%%%%%%%%%%%%%%%%%%%%%%%%%%%%%%%%%%%%%d%%%%%%%%%%%%
\begin{eqnarray}{\label{eq:stfdm_width_decay_to_Pion}}
\Gamma_{\chi^{\pm}} = (n^2 -1) \dfrac{G_{F}^{2}V_{ud}^2 \sin^2(\alpha)  \Delta M^3 f_{\pi}^2 }{4\pi} \sqrt{1 - \dfrac{m_{\pi}^{2}}{\Delta M ^{2}}}~,
\end{eqnarray}
with $n=3$, $G_F$, the Fermi constant, $V_{ud}$, the  up-down element
in the CKM quarks mixing matrix, $f_{\pi}= 131~\rm{MeV}$,
$m_{\pi^{\pm}}=139.570 ~\rm{MeV}$, the charged Pion mass and $\Delta M = m_{\chi^{\pm}} - m_{\pi^{\pm}} $.
In  the CMS analysis~\cite{Sirunyan:2018ldc}, a
search of long-lived charginos in a supersymmetry model
is carried out, 
using disappearing track signatures
and exclude
charginos with lifetimes from $0.5$~ns to $60$~ns for chargino
massese of $505$~GeV. These limits does not apply
directly in this scenario. However,
such an analysis has the potential
to explore the red
points  on the Fig~\ref{fig:stf_collider_1}.
But, this will require a careful treatment 
that is currently beyond the scope of this work.
Nevertheless, using the results from the
CMS analysis mentioned above, 
it is still possible to constrain a
fraction of the parameter space of  scenario {\bf C}.
On the Fig.~\ref{fig:stf_width_cms}, the solid black
line is the {\rm{NLO}} theoretical cross section at
$\sqrt{s}=13$~TeV times branching ratio
for the direct production of a pair of charged fermions,
which latter decay to a charged a Pions and a  fermionic DM specie.
The solid blue, red and green lines stands for the observed $2\sigma$
limits on $\sigma(pp \to \chi^{+} \chi^{-} ) \times Br(\chi^{\pm} \to  \chi_{1}^{0} \pi^{\pm})$
for fermions with lifetimes of  $0.33$~ns, $3.3$~ns,
and $33$~ns, which allow it to exclude  fermion triplets
with masses $m_{\chi^{+}}$  up to
 $320~\rm{GeV}$, $550~\rm{GeV}$ and $380~\rm{GeV}$  respectively.

%%%%%%%%%%%%%%%%%%%%%%%%%%%%%%%%%%%%%%%%%%%%%%%%%%%%%%%%%%%%%%%%  

 \subsection{Benchmarks points}

   Finally, a set of two benchmark scenarios, which fulfills all constraints are given in table
   \ref{Fig:BP_FTM}.

 \begin{table}[b!]
\begin{center}
\begin{tabular}{|l | l | l | l|}

\hline

BP & Scalar Parameters  & Fermion Parameters & Observables \\
\hline

BP1& $\mu_{\eta}^2=7.65\times 10^4 $ GeV$^2$, ~ $\lambda_3=-5.49\times 10^{-3} $      & $m_{\Sigma}=825.28$ GeV   & $\Omega h^2_{\chi_1^0}=1.62 \times 10^{-2}$      \\
& $\lambda_4=-5.33 \times 10^{-3}$, ~ ~ ~ $\lambda_5=-3.21 \times 10^{-3}$         & $m_{N}=1307.98$ GeV & $\Omega h^2_{\eta^R}=1.04 \times 10^{-1}$      \\
& $\lambda_1^{\Omega}=-1.41 \times 10^{-2}$, ~ ~ ~  $\lambda_2^{\Omega}=6.34 \times 10^{-2}$        & $Y_{\Omega}=0.91$     & $\Delta=8.35\times 10^{-2}$     \\
& $\lambda^{\eta}=2.73 \times 10^{-3}$, ~ ~ ~ ~ ~ $\mu_{1}=42.9 $  GeV &  & \\
& $\mu_{2}=15.4 $ GeV &  & \\
\hline
\hline
BP2& $\mu_{\eta}^2=2.52\times 10^5 $ GeV$^2$, ~ $\lambda_3=-1.42\times 10^{-3} $      & $m_{\Sigma}=723.31$ GeV   & $\Omega h^2_{\chi_1^0}=8.73 \times 10^{-2}$      \\
& $\lambda_4=-8.70 \times 10^{-3}$, ~ ~ ~ $\lambda_5=-4.19 \times 10^{-3}$         & $m_{N}=674.03$ GeV & $\Omega h^2_{\eta^R}=3.29 \times 10^{-2}$      \\
& $\lambda_1^{\Omega}=3.23 \times 10^{-3}$, ~ ~ ~ ~ $\lambda_2^{\Omega}=3.18 \times 10^{-3}$        & $Y_{\Omega}=0.69$     & $\Delta=4.31\times 10^{-2}$     \\
& $\lambda^{\eta}=-4.54 \times 10^{-3}$, ~ ~ ~ $\mu_{1}=31.62 $  GeV &  & \\
& $\mu_{2}=15.27 $ GeV &  & \\

\hline
\end{tabular}
\end{center}
\label{Fig:BP_FTM}
\caption{Two benchmark scenarios, for both we fixed  $v_{\Omega}= 5$ GeV and $\lambda_2= 0.2$.  
}
\end{table}

\section{Conclusion}
\label{sec:conclusions}
In this work, we have explored three multicomponent dark matter models with two DM candidates. All models have in common that, in the scalar sector, the DM candidate is the lightest neutral particle of the IDM or an inert scalar, while the other candidate is the lightest neutral mass eigenstate resulting from a mixture of fermionic fields. In this last sector, we focused on a minimal approach, including only fields that are singlets, doublets, and triplets under the $SU(2)_L$ group and allowing them to mix. After spontaneous symmetry breaking, we find that, for all models, the lightest neutral fermionic particle, the DM candidate, is a Majorana fermion. For all models, we imposed theoretical constraints such as those arising from oblique parameters, the Higgs diphoton decay rate, LEP limits, vacuum stability, and perturbativity. Taking this into account, we scanned the available parameter space and study the restrictions resulting from DD, ID and collider experiments. When possible, we also presented future prospects.

For the SDFDM+IDM, we considered a vector-like doublet and a Majorana singlet, the fields mix and, after EWSB the model includes three neutral Majorana particles and one charged fermion. 
The interplay of the two DM candidates can explain the relic abundance for masses from $60$ GeV to the TeV scale. Remarkably, although the region for $100\, \text{GeV} \lesssim m_{H^0} \lesssim 550\, \text{GeV}$ can explain the relic abundance, it is due to the contribution of the fermion field. The DM conversion mechanism does not play an important role when we imposed the current experimental value for the relic abundance. 

Regarding DD, although points with large $\lambda_L$ and large $c_{\chi_1^0\chi_1^0 h}$ generate a huge SI cross section and they could exceed the XENON1T limit, they can not be excluded because they could have a low contribution to the relic density of the DM. The interplay between the relic density, the SI and SD cross section needs to be taken into account as it is shown by the eq.~\ref{eq:delta}.
On the other hand, regarding ID, the thermal averaged annihilation cross-section $\langle \sigma v\rangle$ always falls under the current Fermi-LAT limits for different annihilation channels. However, the model could be tested in future experiments such as CTA. 
Finally, for the case of collider searches, we found that, for the case of the fermionic DM, the spectrum is compressed and, it is hard to put further restrictions on the model. 
   
In the case of the DTF+IDM, we considered a vector-like doublet and a Majorana triplet, the fields mix and, after EWSB the model includes three Majorana fermions and two additional charged ones. Moreover, the scalar sector of the SM is extended with the IDM. The interplay of the two DM candidates allows for the saturation of the relic abundance for the mass of $H_0$ near $80$ GeV, $200 < m_{H_0} < 1200$ GeV, and for $80.0 < m_{\chi_1^0} <1000$ GeV. In the case of DD experiments, XENON1T restricts $|c_{h \chi_1^0 \chi_1^0}|$ to be smaller than $0.08$. On the other hand, current observations from ID experiments place no further restrictions on the parameter space. For the case of collider searches, we found that, in the fermionic sector, due to the mass splittings between the next-to-lightest and lightest fermion, and due to the production cross sections that are mostly doublet, it is hard to put further restrictions on the model. 

For the STFDM model, the scalar DM spice resembles the lightest neutral scalar component of the scalar inert doublet and
the fermion DM candidate arise as the lightest neutral component of the mixing between a $SU(2)$ fermion triplet and a Majorana fermion.
The two DM species can account for the observed relic density in the mass windows $100 < m_{\rm{DM}} <1000$~GeV.

Regarding DD experiment, the XENON1T experiment constrains 
$|\lambda_L| \lesssim {\mathcal{O}}(0.2) $.
In the case of collider searches, the benchmark scenario in which $m_{H^{+}} > m_{\chi^{+}_1} \approx m_{\chi^{0}_2} > (m_{\chi^{0}_1} + m_{W^{\pm}})  $
is explored following  the re-intepretation of an ATLAS analysis, which leads
to the exclusion of  fermion triplets (either $\chi^{+}, \chi^{0}_{2}$ ) with masses of $m_{\chi^{+}} / m_{\chi^{0}_2} \approx 650~\rm{GeV}$
for $m_{\chi_1^{0}}  \approx 200~\rm{GeV}$. And for the compressed mass spectra scenario
 $m_{\pi^{\pm}} <(m_{\chi^{+}} - m_{\chi^{0}_1}) < 0.5~\rm{GeV} $, 
fermions with lifetimes of  $0.33$~ns, $3.3$~ns, and $33$~ns, are excluded  for  fermion triplets with masses $m_{\chi^{+}}$  up to
$320~\rm{GeV}$, $550~\rm{GeV}$ and $380~\rm{GeV}$  respectively. Additionally, current ID 
experiment does not put any restriction on the parameter space of the model.

%AKCNOWLEDGEMENT
\section{Acknowledgment}
We are thankful to Óscar Zapata for useful discussions. G.P and A.B are supported through Universidad EIA grant CI12019007. 
AR is supported by COLCIENCIAS through the ESTANCIAS POSTDOCTORALES program 2017 and Sostenibilidad-UdeA.

%\appendix
\begin{appendix}
\label{ewpo_stfdm}
\section{Oblique parameters in the STFDM model}
\label{ewpo_stfdm}
In the STFDM there are additional contributions to the Peskin-Takeuchi
oblique parameters~\cite{Peskin:1991sw}.
The $S$  and $T$ parameters at one loop level coming from
the scalar sector (inert doublet model plus scalar triplet) and
the singlet-triplet fermion sector are expressed by\footnote{ The U parameters turns out to be small in this kind of models.}:
%%%%%%%%%%%%%%%%%%%%%%%%%%%%%%%%%%%%%%%%
\begin{eqnarray}{\label{eq:STU_stdm_all_1}}
  S_{\rm{new}} &=& S_{\rm{IDM}} + S_{\rm{STM}} + S_{\rm{STF}}~, \\
  T_{\rm{new}} &=& T_{\rm{IDM}} + T_{\rm{STM}} + T_{\rm{STF}}~,    
\end{eqnarray}   
%%%%%%%%%%%%%%%%%%%%%%%%%%%%%%%%%%%%%%%%
where the contribution coming from the IDM reads~\cite{Barbieri:2006dq,Belyaev:2016lok}:
%%%%%%%%%%%%%%%%%%%%%%%%%%%%%%%%%%%%%%%%
\begin{eqnarray}{\label{eq:STU_IDM_stdm_2}}
  S_{\rm{IDM}} &=& \dfrac{1}{72\pi}\dfrac{1}{(x_2^2 -x_1^2)^3} \Bigg (x_2^6 f_a(x_2) - x_1^6 f_a(x_1) + 9x_1^2x_2^2 \Big[ x_2^2f_b(x_2) - x_1^2f_b(x_1)  \Big]  \Bigg )~, \\
  T_{\rm{IDM}} &=& \dfrac{1}{32\pi\alpha^2v^2} \Bigg( f_c(m_{\eta^+}, m_{\eta^{I}}) + f_c(m_{\eta^+}, m_{\eta^{R}}) - f_c(m_{\eta^{R}}, m_{\eta^{I}})  \Bigg)~,
\end{eqnarray}   
%%%%%%%%%%%%%%%%%%%%%%%%%%%%%%%%%%%%%%%%
with $x_1=m_{\eta^{R}}/m_{\eta^{+}}$, $x_2=m_{\eta^{I}}/m_{\eta^{+}}$, $f_a(x)=-5+12\log(x)$, $f_b(x)=3-4\log(x)$ and $f_c(x,y)$
is given by:
%%%%%%%%%%%%%%%%%%%%%%%%%%%%%%%%%%%%%%%%
\begin{align}{\label{eq:STU_stdm_all_3}}
 f_c(x,y)= \left\{ \begin{array}{cc} 
                \frac{x+y}{2} - \frac{xy}{x-y}\log \big (\frac{x}{y} \big ) & \hspace{5mm} x \neq y \\
                0 & \hspace{5mm} x=y \\
                \end{array} \right.
\end{align}
The contribution to $S$ and $T$ arising  from the scalar triplet reads~\cite{Forshaw:2001xq}:
%%%%%%%%%%%%%%%%%%%%%%%%%%%%%%%%%%%%%%%%
\begin{eqnarray}{\label{eq:STU_TM_stdm_4}}
  S_{\rm{STM}} &=& 0~, \\
  T_{\rm{STM}} &=& \dfrac{1}{8\pi}\dfrac{1}{s_{W}^2 c_{W}^2} \Bigg[ \dfrac{m_{h_2}^2 +m_{h^{+}}^2}{m_Z^2} - \dfrac{2m_{h^{+}}^2 m_{h_2}^2}{m_Z^2 (m_{h_2}^{2} -
      m_{h^{+}}^{2})} \log\bigg( \dfrac{m_{h_2}^2}{m_{h^{+}}^2}\bigg) \Bigg]~,
\end{eqnarray}
%%%%%%%%%%%%%%%%%%%%%%%%%%%%%%%%%%%%%%%%
And finally, the contribution to the oblique parameters coming from the singlet-triplet fermion

\begin{eqnarray}{\label{eq:STU_ST_stdm}}
  S_{\rm{STFM}} &=& 0~, \\
  T_{\rm{STFM}} &=&  \dfrac{1}{\alpha} \Bigg ( \dfrac{\Pi_{W W}(0) }{m_{W}^{2}} - \dfrac{\Pi_{Z Z}(0) }{m_{Z}^{2}} \Bigg)~.
\end{eqnarray}

Following the notation in reference~\cite{Cai:2016sjz}, the $\Pi_{V V}$ functions reads:

\begin{eqnarray}{\label{eq:f_self_pi_ww}}
  \Pi_{Z Z}(p^2)  &=& \dfrac{ g_{Z\chi^{+} \chi^{-}}^2}{8\pi^{2}}   \Bigg ( J_1(p^2, m_{\chi^{\pm}}^{2}, m_{\chi^{\pm}}^{2} )
  -2 m_{\chi^{\pm}}^{2} B_0(p^2, m_{\chi^{\pm}}^{2}, m_{\chi^{\pm}}^{2} )    \Bigg )~, \nn \\
  \Pi_{Z Z}(p^2)  &=& \dfrac{1}{8\pi^{2}} \Bigg ( \sum_{i=1}^{2} |a_{W\chi_i^{0} \chi^{\pm}}|^2 \Bigg [ J_1(p^2, m_{\chi_{i}^{0}}^{2}, m_{\chi^{\pm}}^{2} )
  -2 m_{\chi_{i}^{0}} m_{\chi^{\pm}} B_0(p^2, m_{\chi_i^{0}}^{2}, m_{\chi^{\pm}}^{2} )  \Bigg]  \Bigg )~.
\end{eqnarray}
      
where:

\begin{eqnarray}{\label{eq:f_functions_AB}}
  J_1(p^2, m_{1}^{2}, m_{2}^{2}) &=& A_0(m_{1}^{2})  + A_0(m_{2}^{2}) - (p^2 - m_{1}^{2} - m_{2}^{2} )B_0(p^2, m_{1}^{2}, m_{2}^{2})~,
  -4B_{00}(p^2, m_{1}^{2}, m_{2}^{2})~,   \nn \\
  g_{Z\chi^{+} \chi^{-}}&=&g c_W~, \nn \\
  a_{W\chi_1^{0} \chi^{\pm}}&=&g\cos\alpha~, \nn \\
  a_{W\chi_2^{0} \chi^{\pm}}&=&g\sin\alpha~.  
\end{eqnarray}

with $A_0(m^{2})$, $B_0(p^2, m_{1}^{2}, m_{2}^{2})$
and $B_{00}(p^2, m_{1}^{2}, m_{2}^{2})$  Passarino and Veltman scalar integrals~\cite{Passarino:1978jh,Denner:1991kt}.

\end{appendix}

\bibliographystyle{apsrev4-1long}
\bibliography{multicomponent}






% \end{thebibliography}

\end{document}